\documentclass[prl,aps,twocolumn,showpacs]{revtex4}

\usepackage{setspace}
\usepackage{psfrag}
\usepackage{epstopdf}
\usepackage{graphicx}

\usepackage{amsmath,graphicx,epsfig,bm,color}
\usepackage{amssymb,amsfonts}
\usepackage{rotating}
\usepackage{graphics}
\usepackage{setspace}

\begin{document}

\title{\bf The dynamics of quantum vortices in a toroidal trap}
\author{Peter Mason and Natalia G. Berloff}
\affiliation{Department of Applied Mathematics and Theoretical Physics,
University of Cambridge, Wilberforce Road, Cambridge, CB3 0WA, United Kingdom}
\date{\today}

\begin{abstract}
The dynamics of quantum vortices in a two-dimensional annular condensate are considered by numerically simulating the Gross-Pitaevskii equation. Families of solitary wave sequences are reported, both without and with a persistent flow, for various values of interaction strength. It is shown that in the toroidal geometry the dispersion curve of solutions is much richer than in the cases of a semi-infinite channel or uniform condensate studied previously. In particular, the toroidal condensate is found to have states of single vortices at the same position and circulation that move with different velocities. The stability of the solitary wave sequences for the annular condensate without a persistent flow are also investigated by numerically evolving the solutions in time. In addition, the interaction of vortex-vortex pairs and vortex-antivortex pairs is considered and it is demonstrated that the collisions are either elastic or inelastic depending on the magnitude of the angular velocity. The similarities and differences between numerically simulating the Gross-Pitaevskii equation and using a point vortex model for these collisions are elucidated. 
\end{abstract}

\pacs{??}
\maketitle

\section*{I. INTRODUCTION}

A Bose-Einstein condensate (BEC) offers an excellent quantum system to investigate, both experimentally and theoretically. Experimentally, virtually all of the physical parameters can be controlled and thus the dynamics of quantum systems on a macroscopic scale are readily observed. It is a challenge, theoretically, to devise relevant and accurate models to accompany the increasing number of physical phenomena currently being reported. Quantum vortices are one such nonlinear physical phenomena that have been observed in BECs \cite{mat,ram,dutton,a,abo,madison}, and that have been investigated theoretically \cite{pis,fetterRev,cb}. In a BEC, the vortices have quantized circulation and their interactions on an inhomogeneous density background have been studied extensively \cite{mb,rp,anglin,nilsen,lhk}.

Recent experimental interest \cite{racnhp,wnsbda} has shifted to focus
on BECs in toroidal trapping potentials which are now more
experimentally accessible. A BEC in a toroidal trap will form a ring
shaped condensate. The first experimental observation of spontaneous
vortex formation as the alkali gas is cooled through the transition
temperature to create a BEC has been reported in \cite{wnsbda} for
both a toroidal and a harmonic trap. In previous studies vortex
formation through the transition temperature could be attributed to a
number of different factors such as stirring of the condensate by a
laser beam. The conditions under which a ring condensate exists and
the manipulation of a condensate in a toroidal trap has recently been
considered by \cite{jk} and the generation of solitary waves have been
studied in \cite{bbrand}. Experimental interest in
the dynamics of superfluids in a ring is not new. Donnelly \& Fetter
\cite{df66} noted that, in superfluid $^4$He, vortices appear in the
ring at a particular angular velocity when vortices can first
compensate for the difference in irrotational velocity between the
inner and outer boundaries of the ring. Experimentally, rotation of
the condensate will create instabilities around the edge of the
condensate. As the rotation velocity is increased the instabilities
continue to grow and  form vortices (see \cite{tku,cozzini}). A three-dimensional toroidal trapping potential, which creates a ring shaped condensate, might provide the necessary energy requirements to ensure stationary solitary wave solutions exist.

Solitary wave sequences have previously been considered in both two- and three-dimensional homogeneous condensates \cite{jr1982,jpr1986}, a three-dimensional cigar shaped condensate \cite{kp1,kp2,kp3} and also a two-dimensional channel condensate \cite{mb}. The toroidal geometry considered here ensures that there are no phase differences in the condensate. As a result, the dispersion curve of the solitary wave sequences is far richer than has previously been seen. 

In a recent experiment, the observation of persistent flow in a toroidal trap and the existence of vortices, has been reported in \cite{racnhp}. Persistent flow in a BEC is the persistent circulation facilitated by the frictionless flow of a superfluid system. A useful way to view persistent flow is to consider a single vortex pinned to the centre of the ring shaped condensate. The energy at the centre is a (local) minimum and as such it will cost too much energy for the vortex to drift away from the centre. The vortex creates a constant circulation around itself: a persistent flow. A persistent flow in a BEC is also referred to as a supercurrent. The properties of a persistent flow in a toroidal trap have been theoretically considered in \cite{modugno}. It is thought that studies on persistent flows in a BEC held under a toroidal trap could lead to crucial insights into the features of the critical velocity in superfluid $^4$He and to the fundamental relationship between superfluidity and Bose-Einstein condensation \cite{racnhp}.

A two-dimensional condensate with external potential trap of the form used by \cite{racnhp} is considered
\begin{eqnarray}
	\label{3dtor_trap}
	V_{ext}(x,y,z)&=&\frac{1}{2}m\left(\omega_x^2x^2+\omega_y^2y^2\right)+\nonumber\\
	&\phantom{=}&\quad V_0\exp(-2(x^2+y^2)/w_0^2),
\end{eqnarray}
for a constant potential $V_0$, waist $w_0$ (the axial distance from
the laser beam's narrowest point), mass $m$ and
harmonic frequencies components $\{\omega_x,\omega_y\}$ will create the required ring shaped condensate. Reduction to a two-dimensional condensate is equivalent to a tightly trapped, or confined, condensate in the axial coordinate. Such a confinement is experimentally achievable and reduction to two-dimensional condensates has been theoretically considered in \cite{lowdim}. Previous theoretical investigations \cite{zueva,mssss,sssl,lakaniemi} have concentrated solely on solitary wave solutions in the absence of a persistent flow, in particular vortex solutions, in an idealised two-dimensional annular shaped condensate where the density is uniform within the annulus and zero outside of the annulus (a `boxlike' toroidal trap). Their analysis uses established point vortex model techniques from classical hydrodynamics and has shown that the motion of vortices is similar to the motion of classical vortices in an annular shaped domain \cite{saffman}. The trapping potential (\ref{3dtor_trap}) creates an inhomogeneous condensate that is markedly different to a `boxlike' condensate.

This paper will focus solely on the dynamics of solitary wave solutions in a two-dimensional condensate held under a toroidal trapping potential. Section II will introduce the required mathematical formulation for the annular condensate without a persistent flow. The families of solitary wave sequences that exist are then reported in Sect.\ III. In Sect.\ IV, the stability of the solitary wave sequences are numerically investigated and in Sect.\ V the evolution and collisions of vortex-vortex pairs and vortex-antivortex pairs are considered. Section VI adds a persistent flow to the annular condensate and reports the families of solitary wave sequences. Finally, the paper ends with a conclusion, Sect.\ VII.

\section*{II. FORMULATION}

The dynamics of an annular shaped condensate are accurately described by the time-dependent dimensional Gross-Pitaevskii (GP) equation in terms of the macroscopic wave function $\psi=\psi(r,\theta,t)$,
\begin{equation}
	\label{gp_tor2}
	{\rm i}\hbar\frac{\partial\psi}{\partial t}=-\frac{\hbar^2}{2m}\nabla^2\psi-\left(E_v-V_{ext}(r,\theta)-U_0|\psi|^2\right)\psi,
\end{equation}
for two-dimensional external potential trap $V_{ext}(r,\theta)$. The
two-dimensional coupling constant is $U_0$, the mass of a
boson is $m$ and $E_v$ is the chemical potential of the system. The external potential trap is taken to be solely in the radial direction such that
\begin{eqnarray}
	\label{trap_tor}
	V_{ext}(r,\theta)&\equiv& V_{ext}(r)\nonumber\\
	&=&V_0\exp(-2r^2/w_0^2)+\frac{1}{2}m\omega^2r^2,
\end{eqnarray}
for a constant potential $V_0$, waist $w_0$ and frequency $\omega$ ($\omega\equiv\omega_x=\omega_y$, $\omega_z=0$). For small $r$ the exponential term will dominate over the harmonic term and, for a sufficiently tuned $V_0$, a circular hole in the centre of the condensate, an inner boundary, is created where no atoms will be present. Conversely, at larger values of $r$ the harmonic term will dominate causing the condensate to form an outer boundary. Note that the toroidal trap can also be modelled by an external potential trap of the form
\begin{equation}
	\label{mexican}
	V_{ext}(r)=\frac{1}{2}m\omega^2\left(\lambda\frac{r^4}{b_{\perp}}+\sigma r^2\right),
\end{equation}
for $\sigma=-1$ as has been considered by many authors (see for example \cite{cozzini})
for oscillator length $b_{\perp}$ and relative strength of the quartic term $\lambda$. Furthermore, a toroidal trap can be created by using Eq.\ \ref{mexican} with $\sigma=+1$ and placing the condensate into rapid rotation. Above a certain critical angular velocity of rotation, a central hole is created in the condensate resulting in a toroidal geometry (see \cite{spr,ad,fjs,kf}). However only (\ref{trap_tor}) will be used throughout this paper.

Equation (\ref{gp_tor2}) can be non-dimensionalised according to 
\begin{equation}
	\psi\rightarrow\frac{n^{1/2}}{a_{\perp}}\psi,\qquad t\rightarrow\frac{1}{\sqrt{2}\omega}t,\qquad r\rightarrow a_{\perp}r,
\end{equation}
where the transverse oscillator length is $a_{\perp}=(\hbar/\sqrt{2}m\omega)^{1/2}$ and the number density of the ground state is $n=g\hbar^2/(2mU_0)$ where the non-dimensional coupling potential $g$ has been introduced. Thus the non-dimensional GP equation describing the dynamics in the annulus is
\begin{equation}
	\label{gp_tor3}
	2{\rm i}\frac{\partial\psi}{\partial t}=-\nabla^2\psi-\left(\mu-V(r)-g|\psi|^2\right)\psi,
\end{equation}
for chemical potential $\mu=\sqrt{2}E_v/\hbar\omega$, coupling constant $g=\sqrt{2}nU_0/\hbar\omega a_{\perp}^2$ and where the external potential trap is
\begin{equation}
	\label{trap_toroidal}
	V(r)=A\exp(-l^2r^2)+\frac{1}{2}r^2,
\end{equation}
for potential $A=\sqrt{2}V_0/\hbar\omega$ and inverse waist $l=(\sqrt{2}\hbar/m\omega w_0^2)^{1/2}$. Equation (\ref{gp_tor3}) is a four parameter system $\{g, A ,l,\mu\}$ and is subject to the normalisation
\begin{equation}
	\label{tor_norm}
	\int_{\mathcal{V}}|\psi|^2r\,drd\theta=2\pi\int_{\mathcal{V}}|\psi|^2\,rdr=1,
\end{equation}
where $\mathcal{V}$ is the entire spatial domain.

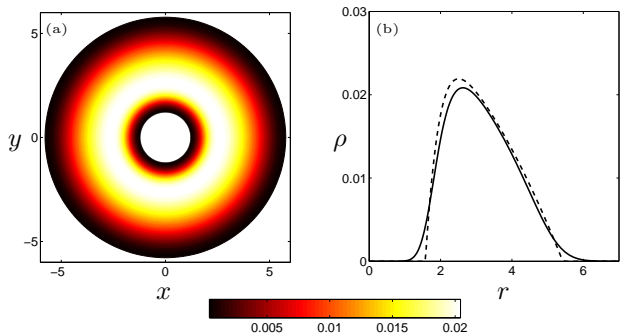
\begin{figure}[ht]
\centering
\includegraphics[scale=0.3]{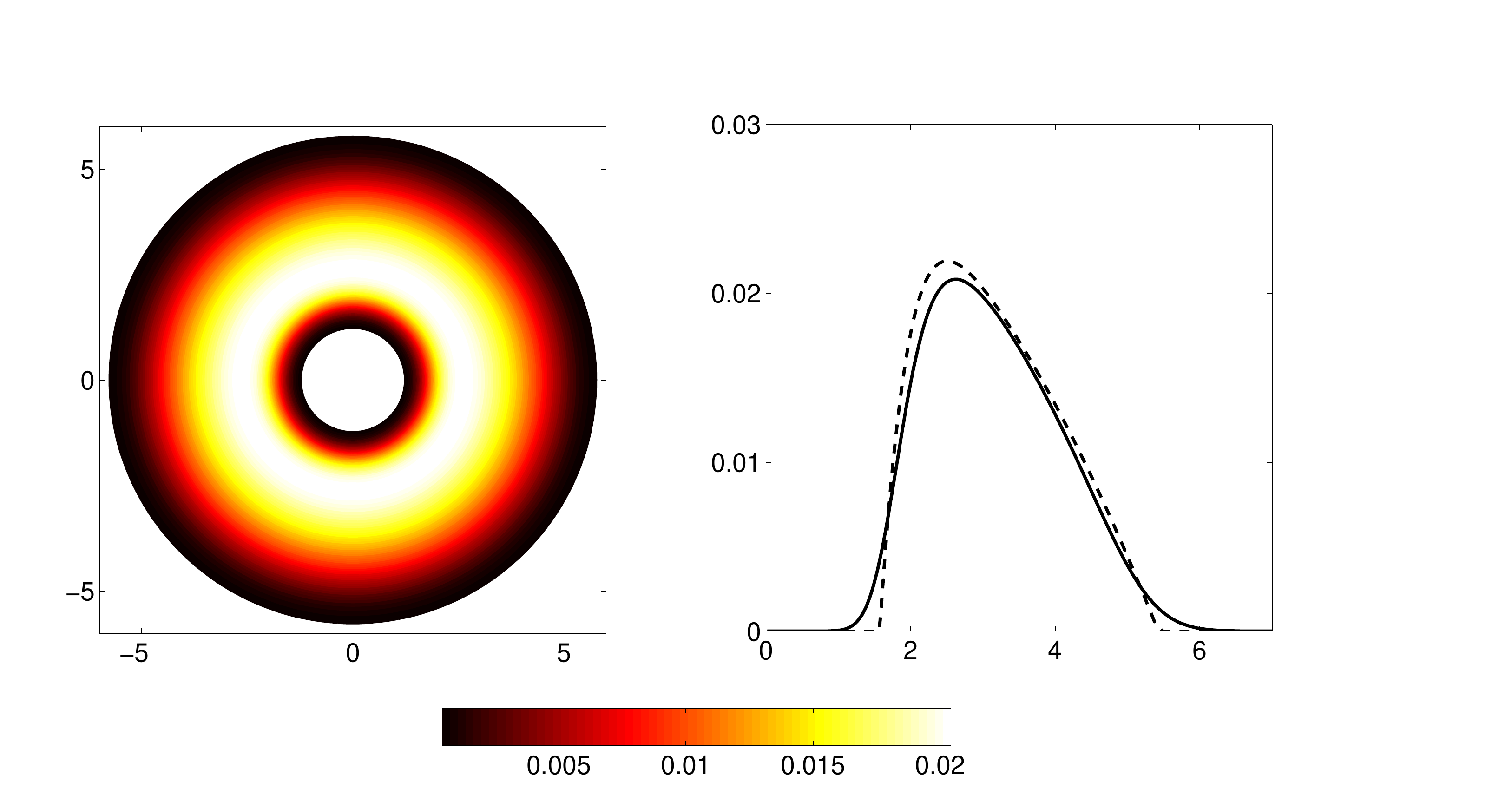}\\
\begin{picture}(0,0)(10,10)
\put(-105,95) {$y$}
\put(-49,37) {$x$}
\put(80,37) {$r$}
\put(18,95) {$\rho$}
\put(-91,138) {{\tiny{(a)}}}
\put(34,138) {{\tiny{(b)}}}
\end{picture}
\vspace{-20pt}
\caption{\baselineskip=-10pt\footnotesize (Color online) The ground states of a condensate acting under a toroidal potential trap (\ref{trap_toroidal}) for $g=500$, $A=100$ and $l=0.9$. Frame (a) is a contour plot of the exact ground state $\psi_0$ while (b) is a slice along constant $\theta$ of the exact ground state $\psi_0$ (solid line) and the Thomas-Fermi approximation $\psi_{\textsc{tf}}$ (\ref{tf_tor}) (dashed line).}
\label{gs_tor}
\end{figure}

A good approximation to the ground state for a particular choice of parameters, $\{g,A,l,\mu\}$, can be garnered from the Thomas-Fermi (TF) approximation, given by 
\begin{equation}
	\label{tf_tor}
	\psi_{\textsc{tf}}(r)=
	\begin{cases}
\left[\frac{\mu-V(r)}{g}\right]^{1/2}, & \mu>V(r),\\
\phantom{xxix}0, & \text{otherwise}.
\end{cases}
\end{equation}
The four parameter set $\{g,A,l,\mu\}$ can be reduced to three by substitution of the TF approximation (\ref{tf_tor}) into the normalisation condition (\ref{tor_norm}), i.e.
\begin{equation}
	2\pi\int_{\mathcal{V}}\frac{\mu-A\exp(-l^2r^2)-\frac{1}{2}r^2}{g}\,rdr=1,
\end{equation}
such that the value of the chemical potential $\mu$ (say) is determined by the choice of $\{g,A, l\}$. An example of the ground state for $g=500$, $A=100$ and $l=0.9$ is shown in Fig.\ \ref{gs_tor}. Note that since the Thomas-Fermi profile (\ref{tf_tor}) is only an approximation to the ground state, a numerical procedure, described in Sect.\ III, is required to find the exact ground state $\psi_0$ for a particular choice of $\{g,A, l\}$. The corresponding chemical potential is found, by the same numerical procedure, to be $\mu=14.72$.

The energy functional and angular momentum for the inhomogeneous condensate are given respectively by \cite{jr1982, mb}
\begin{eqnarray}
	\label{e_tor}
	E_f&=&\frac{1}{2}\int_\mathcal{V}|\nabla\psi|^2+(V(r)-\mu)|\psi|^2+\frac{g}{2}|\psi|^4\, rdrd\theta,\\
	\label{p_tor}
	p&=&\frac{{\rm i}}{2}\int_\mathcal{V}(\psi-\psi_0)\frac{\partial\psi^*}{\partial \theta}-(\psi^*-\psi_0^*)\frac{\partial\psi}{\partial\theta}\, drd\theta.
\end{eqnarray}
The energy of the solitary waves can be determined from $E=E_f-E_g$ where $E_g$ is the energy of the ground state in the absence of any solitary waves and is found numerically by evaluating $E_f$ for $\psi=\psi_0$. Alternatively, an approximate analytically expression for $E_g$ is found by evaluating $E_f$ for $\psi=\psi_{\textsc{tf}}$. 

If solitary wave solutions are found to exist it is entirely plausible
that a multitude of possible solutions could exist for various
$\theta$. A restriction to the type of solutions sought is therefore
taken so that only solutions with density minima along $\theta=\pm\pi/2$ will be considered. Despite this restriction, the essential dynamics of solitary waves in an annulus will still be reflected. 

\section*{III. SOLITARY WAVES}

Solitary wave solutions of Eq.\ (\ref{gp_tor3}) that preserve their
form and move with a constant angular velocity $\Omega$ at constant
$r$ subject to the external potential (\ref{trap_toroidal}) are
sought. A typical example of such a solitary wave solution is a single
vortex, which will move in the inhomogeneous condensate because of the
velocity field created by the surface of the condensate. Other
solitary waves that are sought are a vortex pair, dark (gray and
black) solitons and finite amplitude sound waves -- rarefaction waves. Thus, Eq. (\ref{gp_tor3}) is recast in the frame rotating with the solitary wave using $\theta'=\theta-\Omega t$ so that $\partial/\partial t\rightarrow-\Omega\partial/\partial\theta'$. Hence
\begin{equation}
	\label{gp_tor}
	2{\rm i}\Omega\frac{\partial\psi}{\partial \theta}=\nabla^2\psi+\left(\mu-V(r)-g|\psi|^2\right)\psi,
\end{equation}
where $\theta'$ is replaced by $\theta$ for convenience and is solved subject to boundary conditions
\begin{equation}
\psi(r,\theta,t)\rightarrow\psi_0(r)\qquad\text{as}\qquad r\rightarrow\{0,\infty\}.
\end{equation}
In the above statement for the boundary conditions, the ground state wavefunction at $r=\{0,\infty\}$ is zero, i.e. $\psi(0)=\psi(\infty)=0$.

The solitary wave solutions are found numerically by a Newton-Raphson iteration technique. In order to numerically model the shape of the condensate, a cut in the infinite two-dimensional domain is taken along $\theta=0$ and the domain is unfurled so that the new two-dimensional  rectangular domain in $(r,\theta)$ occupies the upper half plane. This semi-infinite numerical domain is mapped by the transformation $\widehat{r}=\tan^{-1}(Dr)$ to a finite grid $(0,\pi/2)\times(0,2\pi)$ where $D$ is a constant chosen to lie in the range $D\sim0.4-0.8$. The boundary conditions at $\theta=0$ and $\theta=2\pi$ are taken to be periodic. The resulting equations are expressed in second-order finite-difference form. Taking $201\times200$ grid points in the finite domain, the discretised nonlinear equations are solved by a Newton-Raphson iteration procedure using a banded matrix linear solver based on the biconjugate gradient stabilized method. The accuracy of the obtained solutions is verified by evaluating various integral identities. Indeed, an alternative expression for the energy can be found  by taking $\theta\rightarrow b\theta$ for constant $b$ in the expressions for energy and angular momentum (\ref{e_tor}) and (\ref{p_tor}) and considering the variational relationship
\begin{equation}
	\frac{\partial}{\partial b}\left(E-\Omega p\right)\bigg\vert_{b=1}=0.
\end{equation}
The result is an alternative expression for the energy
\begin{equation}
\label{eng_alt}
	E=\int_{\mathcal{V}}\frac{1}{r}\left|\frac{\partial\psi}{\partial\theta}\right|^2\, drd\theta,
\end{equation}
which can be used as a check on the numerical results.

An initial ansatz for the numerical procedure for wave function $\psi$ can be obtained from 
\begin{equation}
	\psi=\psi_0\psi_{sol},
\end{equation}
where $\psi_0$ is the exact ground state wave function (obtained numerically from the TF approximation $\psi_{\textsc{tf}}$) and $\psi_{sol}$ is motivated by the `Tsuzuki' dark soliton solution in homogeneous \cite{tsuzuki} and inhomogeneous \cite{mb,kp2} condensates
\begin{equation}
	\label{tzu_tor}
	\psi_{sol}={\rm i}c_1+c_2\tanh(c_3r\cos\theta),
\end{equation}
for constants $c_1$, $c_2$ and $c_3$ to be chosen. Notice that (\ref{tzu_tor}) creates a soliton at $\cos\theta=0$ to ensure that solitary wave solutions exist only along $\theta=\pm\pi/2$. An alternative ansatz which looks directly for vortex solutions can be found from
\begin{equation}
	\label{tor_ansatz}
	\psi=\psi_0\psi_v,
\end{equation}
where the wave function for the vortex at $(r_0,\pi)$ is
\begin{equation}
\label{ansatz}
	\psi_v=\frac{r\cos\theta+{\rm{i}}(r\sin\theta-r_0)}{\sqrt{r^2\cos^2\theta+(r\sin\theta-r_0)^2+2/\mu}}.
\end{equation}
Additional vortices can easily be included into (\ref{tor_ansatz}) by writing $\psi_v=\prod_{i=1}^N\psi_{i}$, where $N$ is the number of vortices, as required. The position of the vortex is detected numerically by finding zeros of the real and imaginary parts of the wave function. 

As a vortex moves close to the boundary of the
condensate, where $\psi \rightarrow 0$, it becomes impossible to
distinguish between a rarefaction wave and a vortex solution. In what
follows the solitary wave solution is referred to as a ``rarefaction wave'' when the numerics cannot resolve
whether or not the real (imaginary) part of $\psi$ changes its sign
along the polar angle (see also \cite{mb} for discussion).

An asymptotic expression for the group angular velocity is obtained from the variation $\psi\rightarrow\psi+\delta\psi$ in (\ref{e_tor}) and (\ref{p_tor}) and allows a comparison with the results of the numerics to be made. On use of the GP equation (\ref{gp_tor}) one obtains
\begin{equation}
\label{grp}
	\Omega=\frac{\partial E}{\partial p},
\end{equation}
with the derivatives taken along the direction of propagation of the solitary waves. To obtain an approximation to the angular velocity $\Omega$ as a function of $r_0$, the position of the vortex, Eq.\ (\ref{grp}) is rewritten as
\begin{equation}
\Omega=\frac{\partial E}{\partial p}=\frac{\partial E/\partial r_0}{\partial p/\partial r_0}
\end{equation}
for the energy and angular momentum given as in (\ref{eng_alt}) and
(\ref{p_tor}), respectively.

Of the three parameters contained within the system $\{g, A, l\}$ (the value of $\mu$ is determined from the other three), the interaction strength $g$ can be considered to be the dominant factor. The interaction strength directly determines the number of healing lengths that span the condensate and thus in turn directly determines the number of solitary wave solutions, along a constant $\theta$, that can be present in the condensate. A range of values of $g$ have been simulated ($30<g<1000$) of which two typical examples will be detailed below. In each example the energy-angular momentum dispersion curve is symmetric about $p=0$. The effect of this symmetry is to simply consider solitary wave solutions that move with the opposite circulation. In order to aid the analysis throughout this section different shades of colors will be used. Shades of black will represent a condensate with one solitary wave solution present, shades of dark gray (red) will represent a condensate with two solitary wave solutions present and shades of light gray (green) will represent a condensate with either three or four solitary wave solutions present.

\subsection*{A. Low Interaction Strength}
\label{toroidal_results_150}

\begin{figure}[ht]
\centering
\includegraphics[scale=0.3]{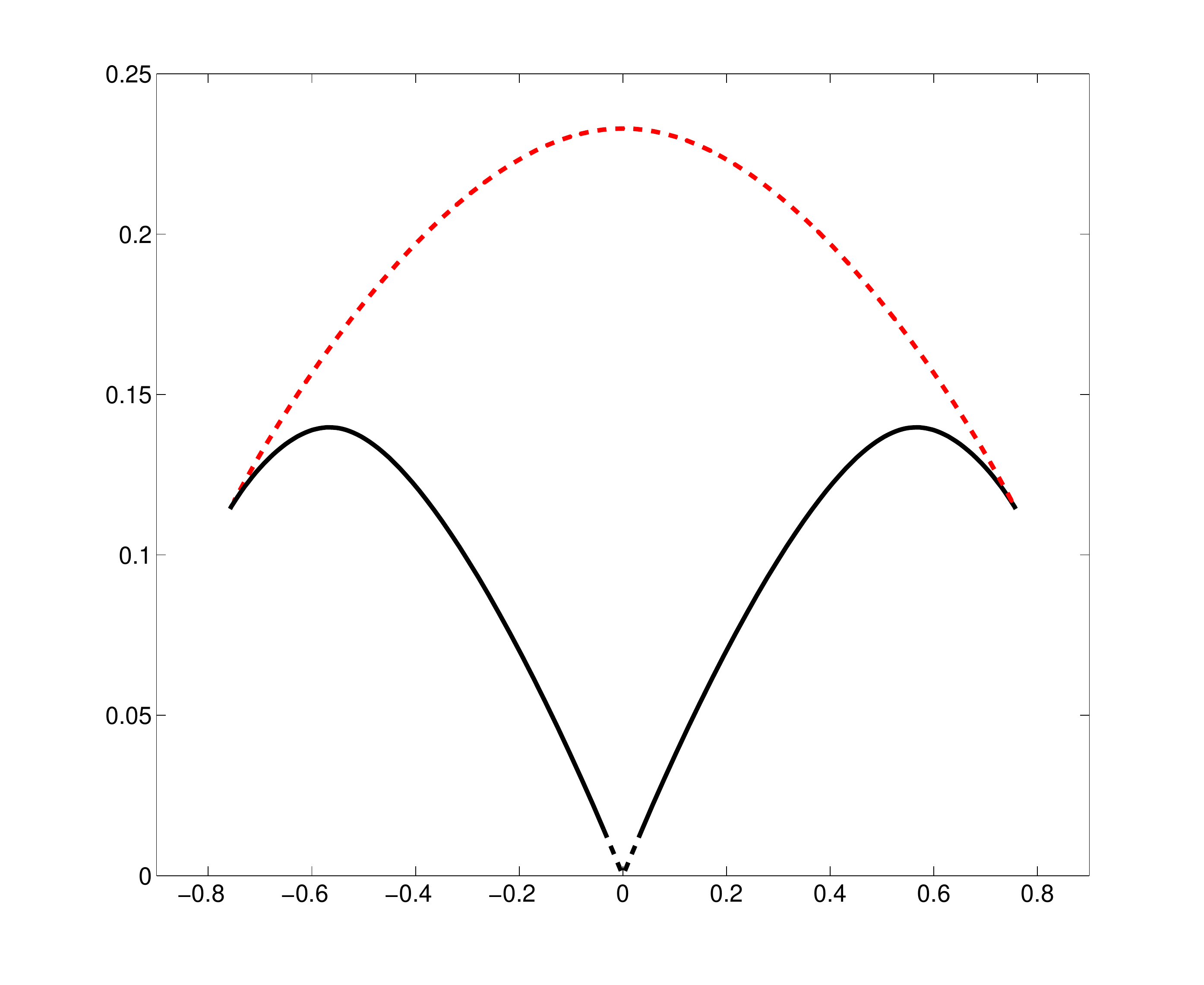}\\
\begin{picture}(0,0)(10,10)
\put(11,26) {$p$}
\put(-102,121) {$E$}
\end{picture}
\caption{\baselineskip=10pt \footnotesize (Color online) The energy-angular momentum dispersion curve for the annular condensate governed by the GP equation (\ref{gp_tor}) for parameter set $\{g, A, l\}=\{150, 40, 1\}$. A single vortex solution for $\theta=\pi/2$ is represented by the thick solid black line while the thick dashed black line represents a rarefaction wave for $\theta=\pi/2$. The thick dashed dark gray (red)  line corresponds to a rarefaction wave for $\theta=\pm\pi/2$. The dispersion curve is symmetric about $p=0$.}
\label{tor_150_ep}
\end{figure}
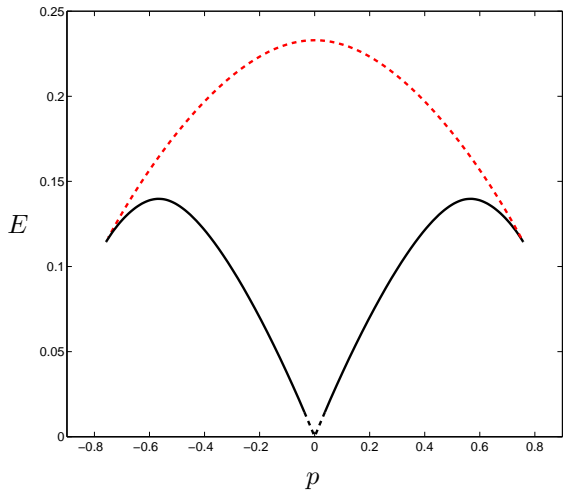

A condensate with a low interaction strength will first be considered with the values of the parameters taken as $\{g, A, l\}=\{150, 40, 1\}$ with the corresponding value of the chemical potential $\mu=8.59$. The complete family of solitary wave solutions in the annulus are then found and the energy-angular momentum dispersion curve is shown in Fig.\ \ref{tor_150_ep}. Two distinct solutions are found to exist. The first is a single vortex (thick solid black line) of positive circulation ($p>0$) which exists for angular velocity $-0.29\le\Omega\le0.40$ and along $\theta=\pi/2$ \cite{num}. The vortex will transcribe circles around the origin for constant angular velocity and for constant radius $r_0$. At $\Omega=0.38$, the single vortex loses its circulation and becomes a rarefaction wave which is present for increasing angular velocities until the termination angular velocity is reached (for the range of parameters chosen here, the termination angular velocity is $c=0.40$). No solutions are present above the termination angular velocity. 

\begin{figure}[ht]
\centering
\includegraphics[scale=0.3]{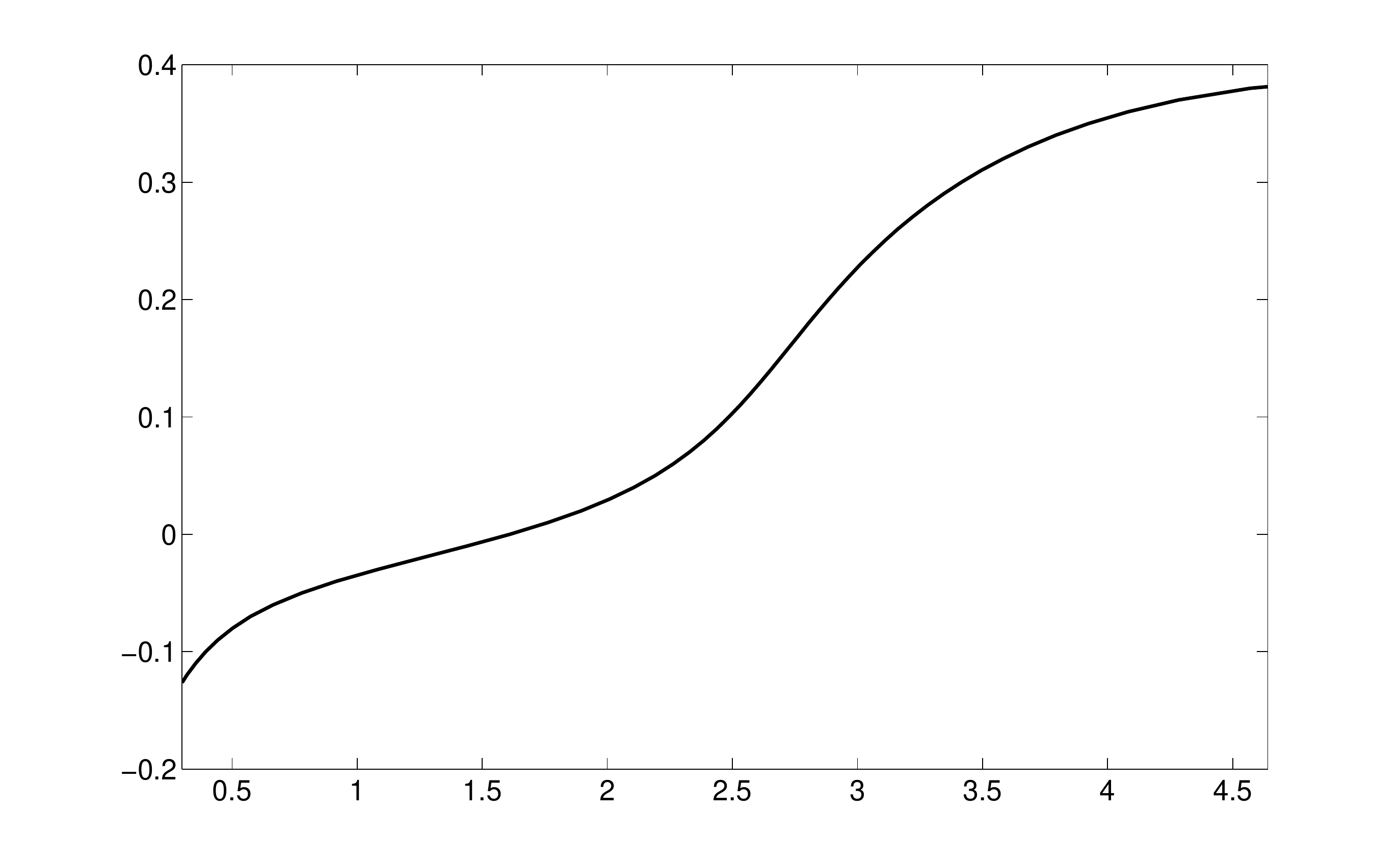}\\
\begin{picture}(0,0)(10,10)
\put(12,22) {$r_0$}
\put(-100,93) {$\Omega$}
\end{picture}
\caption{\baselineskip=10pt \footnotesize The angular velocity $\Omega$ plotted against the distance $r_0$ of the single vortex of positive circulation depicted in Fig.\ \ref{tor_150_ep} for $p>0$. The inner edge of the condensate is at $r=R_1=0.78$ and the outer edge of the condensate is at $r=R_2=4.65$.}
\label{tor_150_ud}
\end{figure}
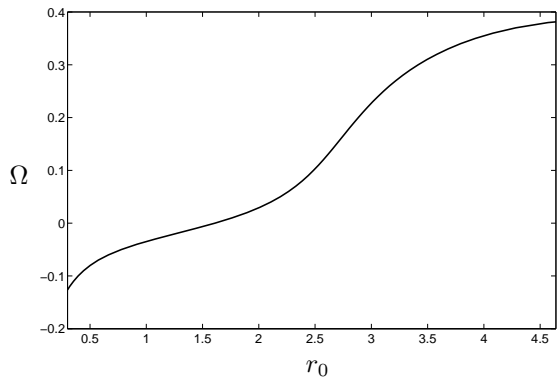

A plot of the angular velocity of the vortex against distance, $r_0$,
from the centre of the condensate is shown in Fig.\
\ref{tor_150_ud}. The shape of the curve is qualitatively similar (in the bulk of the condensate) to
that observed in a `boxlike' toroidal condensate (see \cite{zueva} and
\cite{sssl}). However, at the boundaries of the condensate there is no qualitative agreement: the curve of Fig.\ \ref{tor_150_ud} reaches finite values for the angular velocity at the inner and outer boundaries. Notice that the vortex has zero angular velocity when $r_0\approx1.61$. 

For $\Omega<-0.29$ the single vortex solution does not exist. Instead a cusp is formed similar to that observed in semi-infinite cigar-shaped condensates \cite{kp2,mb}. Over the cusp, where now the angular velocity is increased, a second solution is found to exist as two rarefaction waves along $\theta=\pm\pi/2$ respectively (thick dashed dark gray (red) line in Fig. \ref{tor_150_ep}). The two rarefaction wave solution exists all the way to $p=0$ where a maximum on the dispersion curve is attained and the solution corresponds to a black soliton.

\begin{figure}[ht]
\centering
\includegraphics[width=15mm,bb=300 00 450 500]{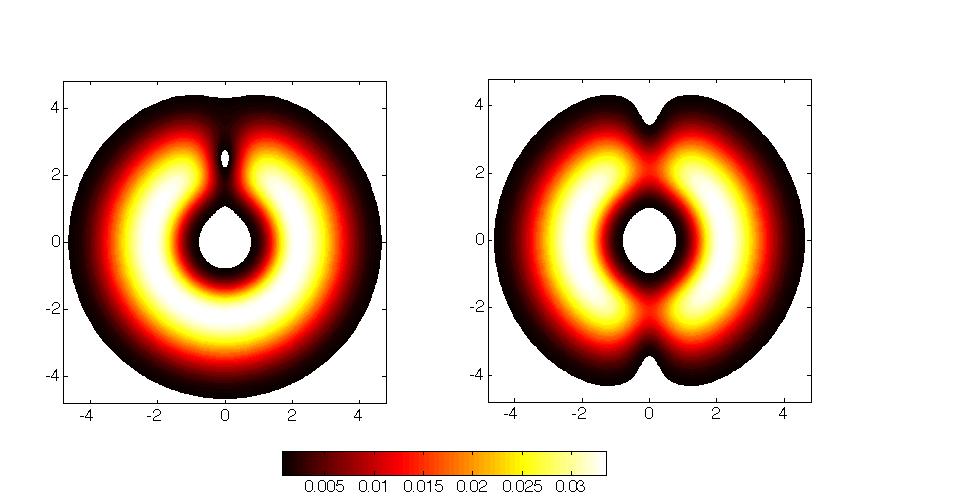}\\
\begin{picture}(0,0)(10,10)
\put(-95,95) {$y$}
\put(-34,37) {$x$}
\put(85,37) {$x$}
\put(-75,134) {{\tiny{(a)}}}
\put(44,134) {{\tiny{(b)}}}
\end{picture}
\caption{\baselineskip=10pt \footnotesize (Color online) Two contour plots depicting two solutions of the GP equation (\ref{gp_tor}) for parameter set $\{g, A, l\}=\{150,40,1\}$. Frame (a) contains a single vortex at $\Omega=0.1$ ($r_0=2.49$) and frame (b) pair of rarefaction waves along $\theta=\pm\pi/2$ for $\Omega=0.2$.}
\label{tor_150_cont}
\end{figure}
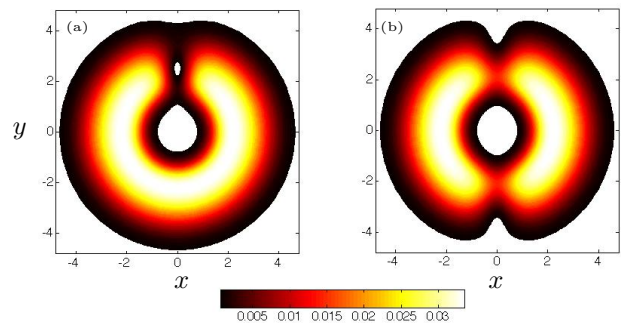

A sample of the two distinct solutions on the dispersion curve is provided by the contour plots in Fig.\ \ref{tor_150_cont}. The single vortex of positive circulation at $\Omega=0.1$ (frame (a)) and the two rarefaction waves at $\Omega=0.2$ (frame (b)) are depicted. Note that by (\ref{grp}), the angular velocity is the gradient of the energy-angular momentum dispersion curve.

\subsection*{B. Mid-Range Interaction Strength}
\label{toroidal_results_350}

\begin{figure}[ht]
\centering
\includegraphics[scale=0.3]{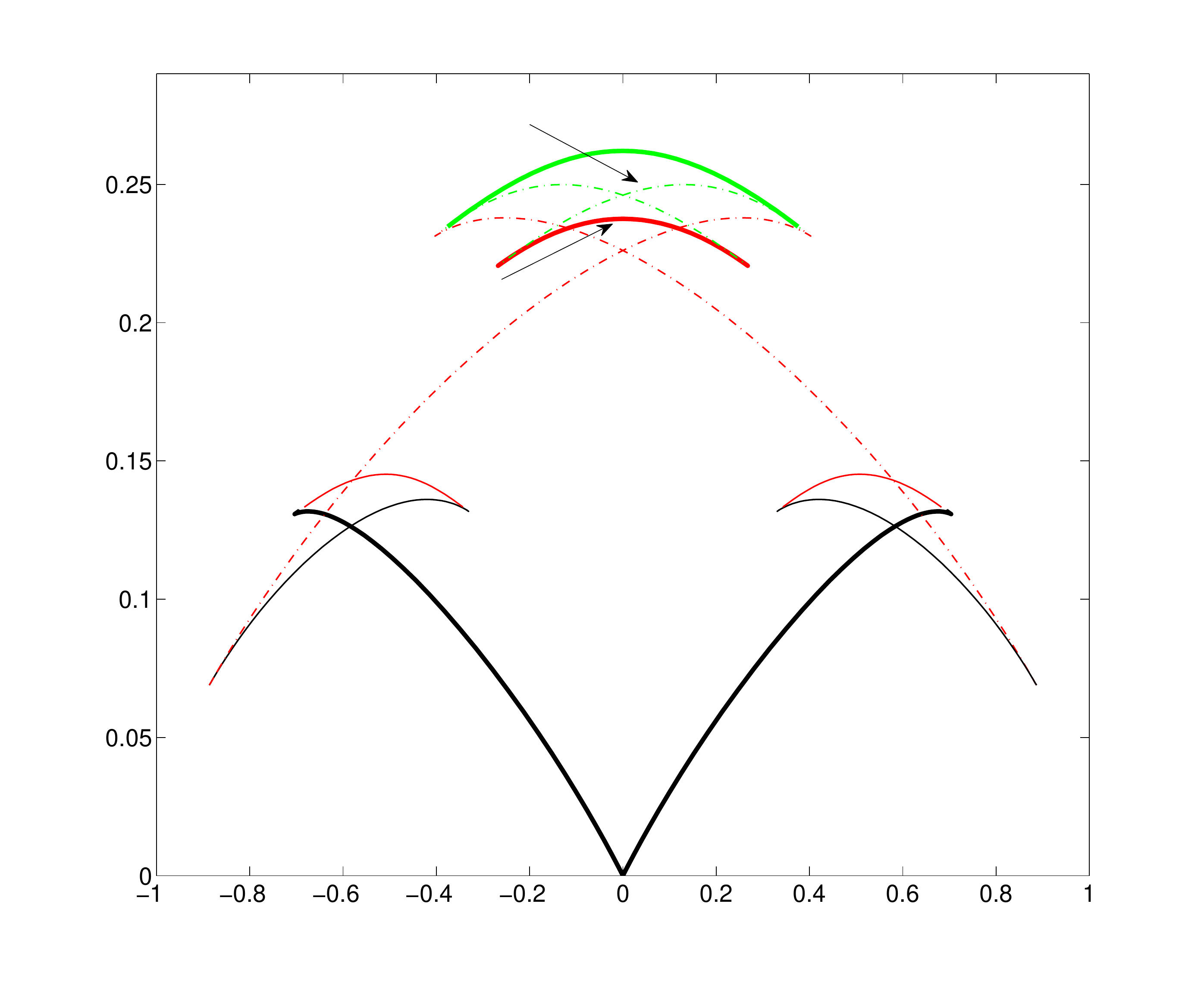}\\
\begin{picture}(0,0)(10,10)
\put(11,26) {$p$}
\put(-102,121) {$E$}
\put(50,92) {{\tiny{(i)}}}
\put(48,126) {{\tiny{(ii)}}}
\put(50,114) {{\tiny{(iii)}}}
\put(50,145) {{\tiny{(iv)}}}
\put(-23,160) {{\tiny{(vii)}}}
\put(18,193) {{\tiny{(v)}}}
\put(-17,197) {{\tiny{(vi)}}}
\end{picture}
\caption{\baselineskip=10pt \footnotesize (Color online) The energy-angular momentum dispersion curve for the annular condensate governed by the GP equation (\ref{gp_tor}) for parameter set $\{g, A, l\}=\{500, 100, 0.9\}$. There are seven distinct branches (i)-(vii) (the others can be obtained by symmetry about $p=0$). Branch (i) corresponds to a single vortex of positive circulation (thick solid black line), branch (ii) corresponds to two vortices on $\theta=\pi/2$ (thin solid dark gray (red) line), branch (iii) corresponds to a single vortex of negative circulation (thin solid black line), branch (iv) corresponds to two vortices at symmetric distances on $\theta=\pm\pi/2$ respectively (thin dashed-dot dark gray (red) line), branch (v) corresponds to the four vortex solution (thick solid light gray (green) line), branch (vi) to a three vortex solution (thin dashed-dot light gray (green) line) and branch (vii) corresponds to two vortices at antisymmetric distances on $\theta=\pm\pi/2$ respectively (thick solid dark gray (red) line).}
\label{tor_500_ep}
\end{figure}
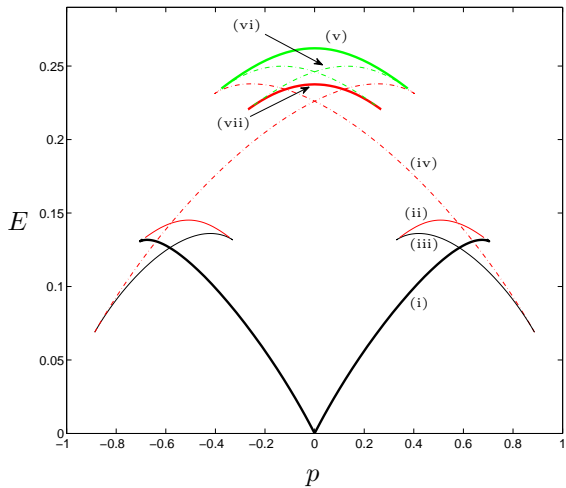

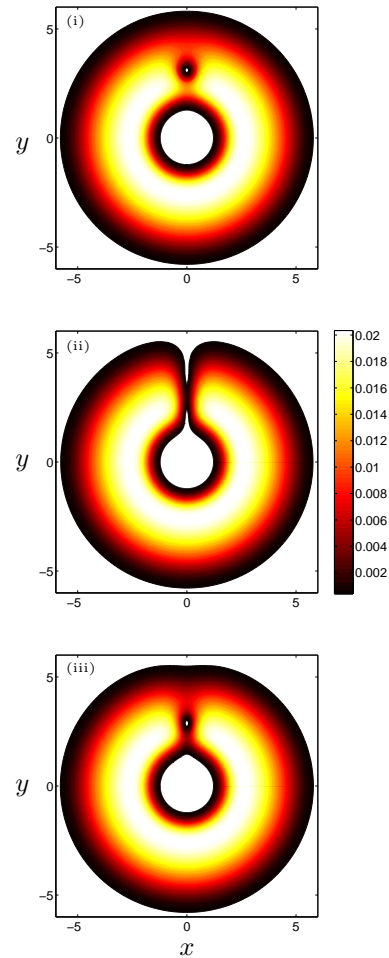
\begin{figure}[ht]
\centering
\includegraphics[scale=0.3]{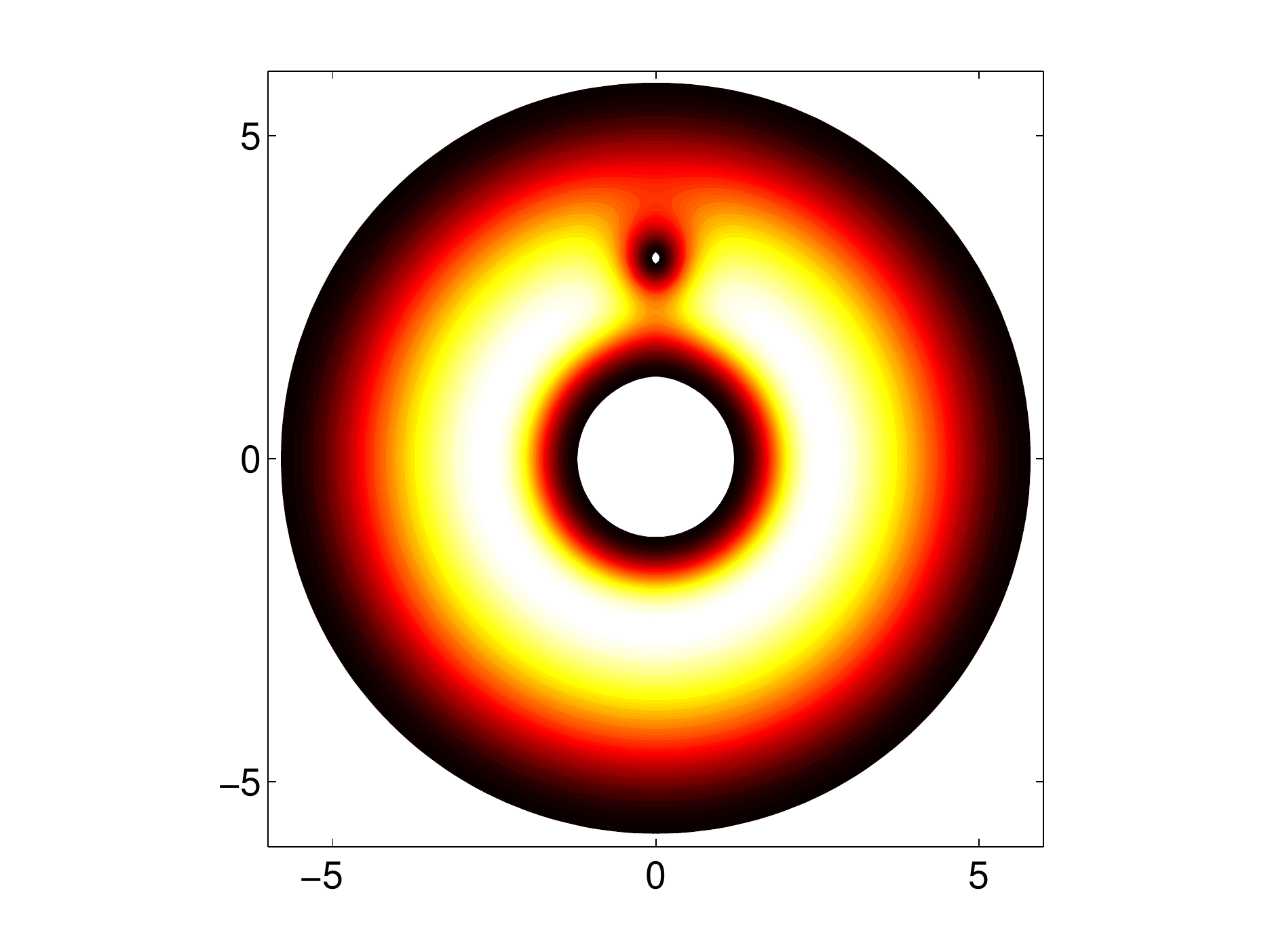}\\
\includegraphics[scale=0.3]{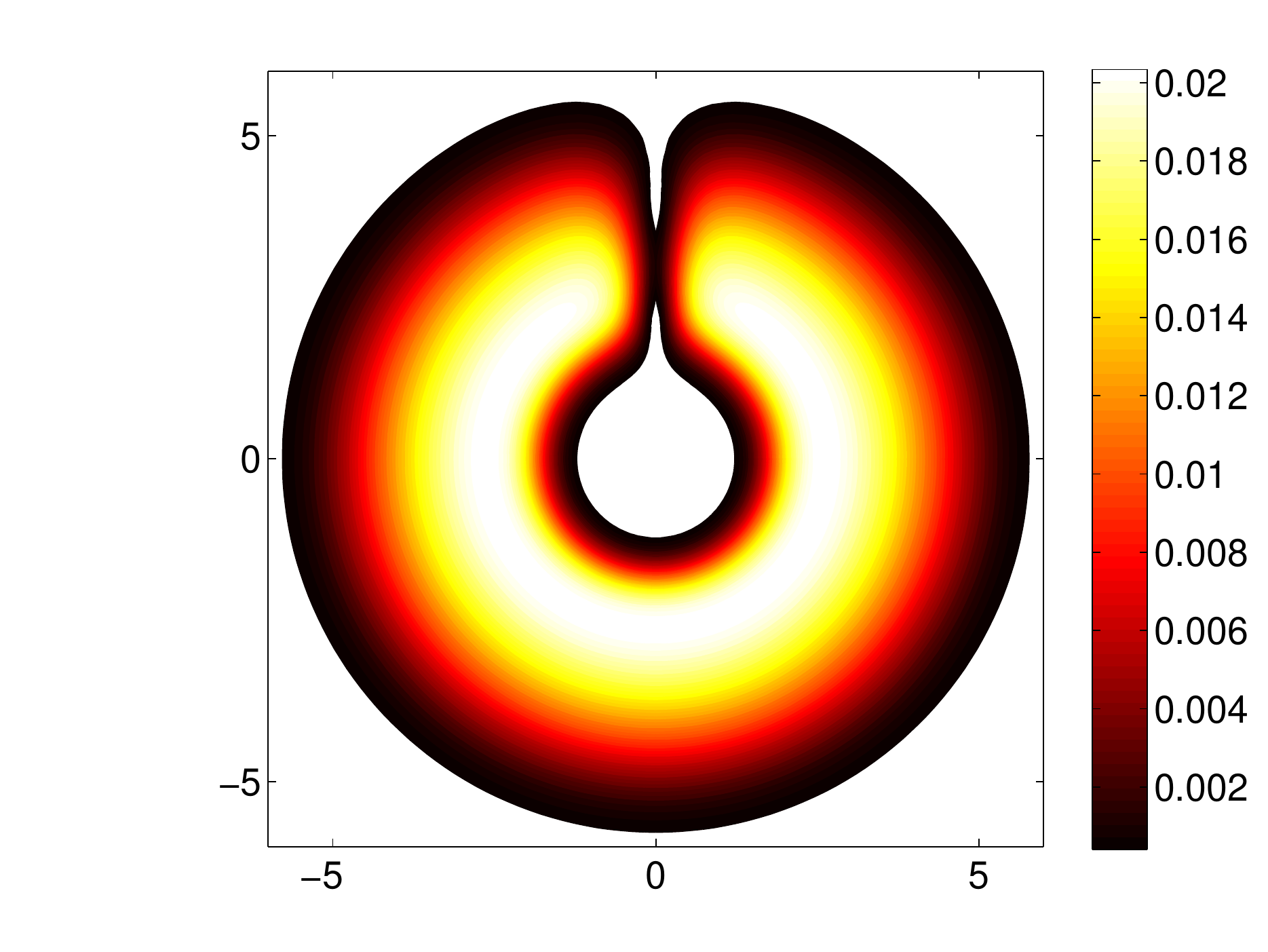}\\
\includegraphics[scale=0.3]{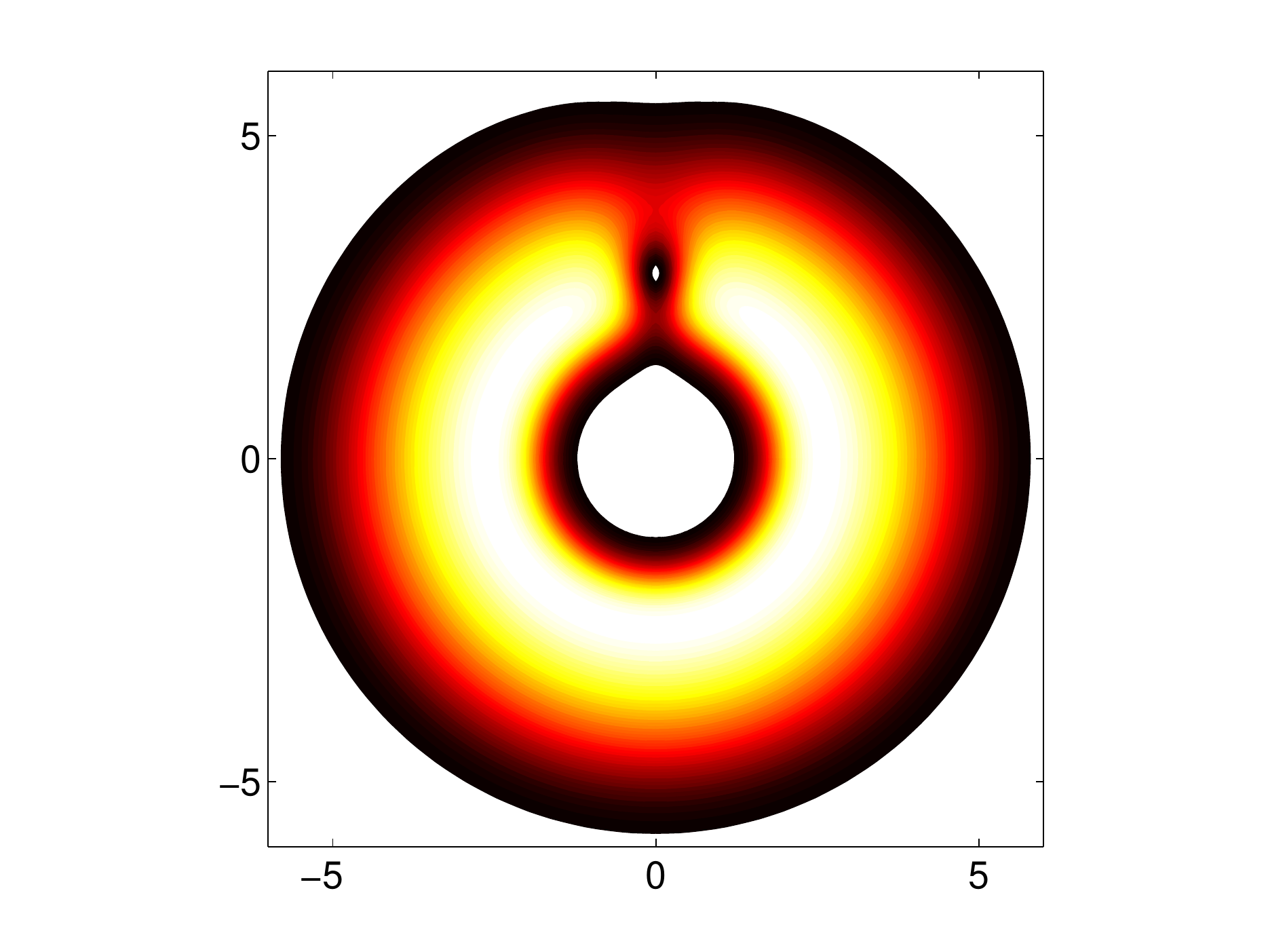}\\
\begin{picture}(0,0)(10,10)
\put(-33,373) {{\tiny{(i)}}}
\put(-33,250) {{\tiny{(ii)}}}
\put(-33,128) {{\tiny{(iii)}}}
\put(-52,326) {$y$}
\put(-52,206) {$y$}
\put(-52,83) {$y$}
\put(10,21) {$x$}
\end{picture}
\caption{\baselineskip=10pt \footnotesize (Color online) Contour plots of the three different configurations (i)-(iii) given on the dispersion curve (Fig.\ \ref{tor_500_ep}) for angular velocities $\Omega=0.1$, $\Omega=0$ and $\Omega=0.1$ respectively.}
\label{tor_350_cont1}
\end{figure}

\begin{figure}[ht]
\centering
\includegraphics[width=100mm,bb=30 00 700 700]{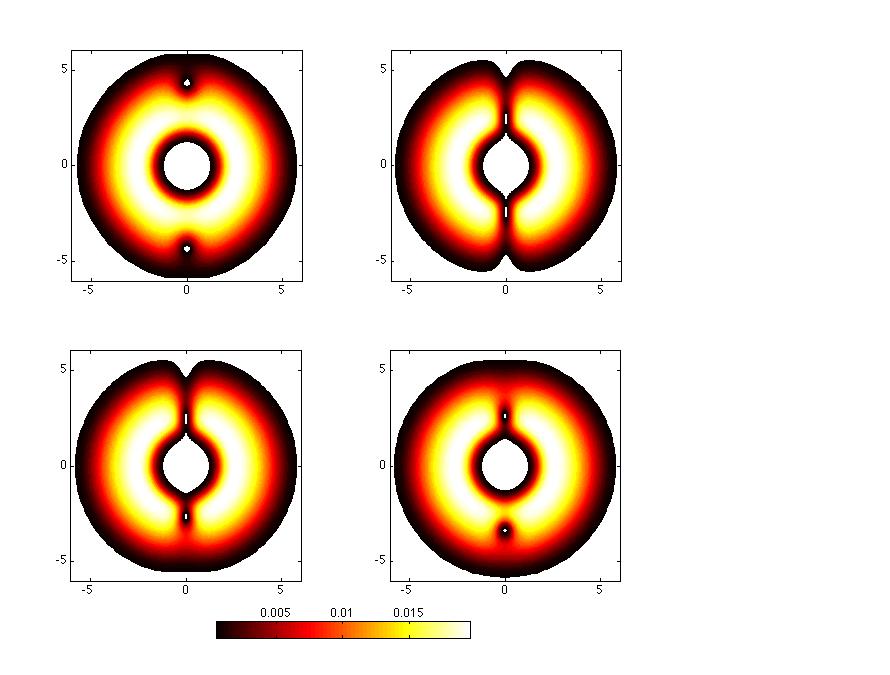}\\
\begin{picture}(0,0)(10,10)
\put(-90,279) {{\tiny{(iv)}}}
\put(42,279) {{\tiny{(v)}}}
\put(-90,152) {{\tiny{(vi)}}}
\put(42,152) {{\tiny{(vii)}}}
\put(-110,230) {$y$}
\put(-110,107) {$y$}
\put(-50,43) {$x$}
\put(84,43) {$x$}
\end{picture}
\caption{\baselineskip=10pt \footnotesize (Color online) Contour plots of the four different configurations (iv)-(vii) given on the dispersion curve (Fig. \ref{tor_500_ep}) for angular velocities $\Omega=0.2$, $\Omega=0.1$, $\Omega=0.1$ and $\Omega=0.1$ respectively.}
\label{tor_350_cont2}
\end{figure}
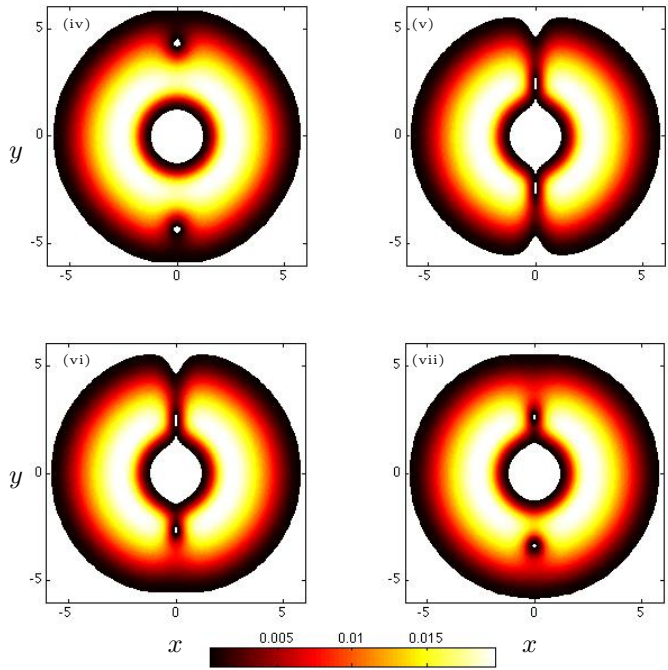

As the interaction strength is increased new features are seen to develop in the energy-angular momentum dispersion curve that are unexpected. An example of the features observed to occur can be seen by taking the parameter set $\{g, A, l\}=\{500, 100, 0.9\}$ with the corresponding value of chemical potential $\mu=14.72$. The dispersion curve is given in Fig.\ \ref{tor_500_ep} and a selection of contour plots for the different solutions of the dispersion curve are given in Fig.'s\ \ref{tor_350_cont1} and \ref{tor_350_cont2}. Note that the dynamics within this particular system are much more diverse than for the low interaction strength $g=150$ detailed in the previous section. The reason for the increase in the number and type of solutions realised is the increase in the condensates' spatial extent. It is now possible for the condensate to contain at least two solitary wave solutions for any $\theta=\pm\pi/2$. 

The dispersion curve contains seven distinct branches with each branch allocated a label (i)-(vii). At zero energy and zero angular momentum there is no solitary wave solution, however an increase in energy and (positive) angular momentum will create a single solitary wave solution moving near the termination angular velocity at the far edge of the condensate along $\theta=\pi/2$. The solitary wave solution is a single vortex of positive circulation. As the angular velocity is decreased, branch (i) in the dispersion curve is transcribed and the position of the vortex is seen to move towards the centre of the condensate. At $\Omega=-0.11$ a kink develops in the single vortex solution branch with energy and angular momentum given by $(E,p)=(0.13,0.70)$. The same single vortex of positive circulation solution is found to be continuous over the kink. The solution is present until $\Omega=-0.12$ at which point the branch terminates with the vortex at $r_0=2.35$. 

A new solution is found to exist that joins with branch (i). The new branch, (ii), is characterised by the creation of a new vortex of negative circulation at the far edge of the condensate on $\theta=\pi/2$; see Fig.\ \ref{tor_350_cont1}, frame (ii) for a contour plot of the new solution. Branch (ii) exists for $-0.12\le\Omega\le0.13$ and includes a solution where $\Omega=0$. As the angular velocity increases both the vortex of positive circulation (near the inner edge) and the vortex of negative circulation (near the outer edge) move towards the origin. At $\Omega=0.13$, where the branch terminates, the positive vortex reaches the inner edge of the condensate and is lost while the negative vortex is at $r_0=2.77$. 

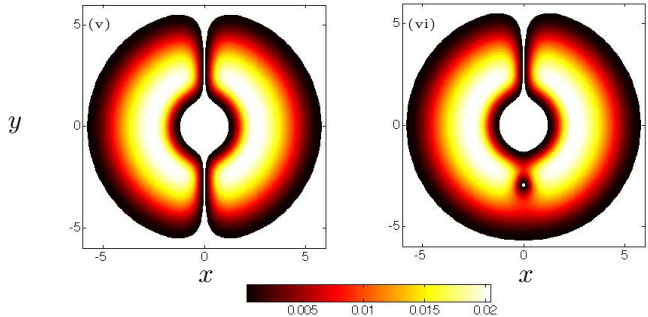
\begin{figure}[ht]
\centering
\includegraphics[width=15mm,bb=300 00 450 500]{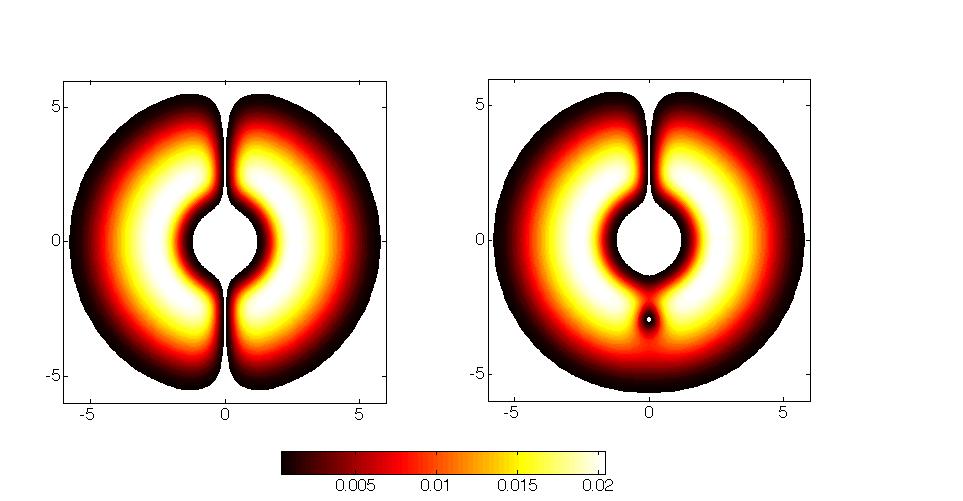}\\
\begin{picture}(0,0)(10,10)
\put(-107,95) {$y$}
\put(-35,37) {$x$}
\put(86,37) {$x$}
\put(-77,133) {{\tiny{(v)}}}
\put(44,133) {{\tiny{(vi)}}}
\end{picture}
\caption{\baselineskip=10pt \footnotesize (Color online) Contour plots for $\Omega=0$ for branches (v) and (vi) on the dispersion curve (Fig. \ref{tor_500_ep}) showing the occurrence of a black soliton.}
\label{tor_350_black}
\end{figure}

A new cusp is formed on the dispersion curve at $(E,p)=(0.13,0.34)$. As the angular velocity is decreased a new branch is traced out (branch (iii)) which is characterised by the existence of a single vortex of negative circulation. Further decreasing the angular velocity shifts the single vortex towards the outer edge of the condensate until at $\Omega=-0.30$ the vortex is lost and the branch terminates.

A further solution can be found by increasing the angular velocity and creating a new branch, (iv). Branch (iv) is characterised by the appearance of two identical vortices at the far edge of the condensate on $\theta=\pm\pi/2$ respectively; see Fig.\ \ref{tor_350_cont2}, frame (iv). The cusp formed by branches (iii) and (iv) is unique to all the other cusps
in the dispersion curve. At the cusp the dynamics contain no solitary wave solutions, similar to the termination point of branch (i) at the termination speed ($p=0$). In the experimental procedures it is possible that disturbances caused outside the far edge of the condensate could theoretically manifest into forming solutions exhibited by branches (i), (iii) or (iv).

The angular velocity range of existence of branch (iv) is $-0.30\le\Omega\le0.13$ during which the energy and angular momentum change considerably. As the angular velocity is increased the two vortices move towards the inner edge of the condensate each keeping the same fixed distance from the origin. The solution can be viewed as a symmetric pair of vortices on $\theta=\pm\pi/2$ (there is symmetry about $y=0$). The solution is present to negative angular momentum and terminates when $(E,p)=(0.23,-0.38)$ with the vortices at $r_0=2.54$.

At this point the dispersion curve forms a three-way cusp. Choosing to follow the thick solid light gray (green) curve in Fig.\ \ref{tor_500_ep} (branch (v)) and thus decreasing the angular velocity, results in two new vortices being created at the edge of the condensate on $\theta=\pm\pi/2$, respectively, to produce a four-vortex configuration; see Fig.\ \ref{tor_350_cont2}, frame (v). The two new vortices are of opposite circulation compared to the two existing vortices. As the angular velocity is decreased towards zero the four vortices all move towards the inner edge of the condensate (the positions of the vortices on $\theta=\pi/2$ always remain symmetric to those on $\theta=-\pi/2$). When the angular velocity becomes zero, two black solitons occur; see Fig.\ \ref{tor_350_black}, frame (v). For negative velocities the dynamics are symmetric to those for positive velocities such that when the branch terminates at $\Omega=-0.13$ for $(E,p)=(0.23,0.38)$, the configuration is identical to that at $(E,p)=(0.23,-0.38)$.

Starting at $(E,p)=(0.23,0.38)$ and instead choosing to follow the thin dashed-dot light gray (green)  curve in Fig.\ \ref{tor_500_ep}, and thus tracing out branch (vi), the penultimate configuration is realised. Here the dynamics consists of three vortices, with the new vortex appearing at the edge of the condensate for $\theta=\pi/2$ (say). Thus the configuration consists of two vortices on $\theta=\pi/2$ and a single vortex on $\theta=-\pi/2$; see the contour plot of Fig.\ \ref{tor_350_cont2}, frame (vi). Increasing the angular velocity will cause all the vortices to gradually move towards the centre of the condensate until at zero angular velocity when the two vortices on $\theta=\pi/2$ decay to form a black soliton while the single vortex on $\theta=-\pi/2$ remains; see Fig.\ \ref{tor_350_black}, frame (vi). For positive velocities the single vortex continues to move towards the centre of the condensate, but the other two vortices now move towards the outer edge of the condensate. As the branch terminates at $\Omega=0.13$, the furthest of these two vortices reaches the outer edge of the condensate and disappears. Meanwhile the two remaining vortices (on opposite sides of the condensate) are at different radii.

The final realisable configuration is mapped out by branch (vii); see Fig.\ \ref{tor_350_cont2}, frame (vii). The two vortices can be described as an anti-symmetric pair since they are at different radii ($r_1=2.61$, $r_2=3.38$ for the vortices on $\theta=+\pi/2$ and $\theta=-\pi/2$ respectively for $\Omega=0.1$) and their dynamics behave differently. As the angular velocity is decreased, the vortex at $r_1$ will move towards the inner edge of the condensate while the vortex at $r_2$ will move towards the outer edge of the condensate. At zero angular velocity $r_1=r_2$ (note that there is no black soliton here). 

\subsection*{C. The Effect of $A$ and $l$}
\label{toroidal_results_Al}

This section will consider in more detail the typical effect that changing the values of the parameters $A$ and $l$ has on the dynamics. As explained before, the interaction strength $g$ is the parameter that is most responsible for the width of the condensate and is thus the crucial parameter determining the type of dynamics that can be exhibited in the condensate. As a result it is thus expected that a change in $A$ and $l$ will not change the number of solitary waves that can span the condensate at any one time (since changing $A$ and $l$ is expected to only have a minor effect on the width of the condensate, and thus the number of healing lengths that span the condensate). 

Thus consider the potential trap (\ref{trap_toroidal}). Taking the derivative with respect to $r$ and searching for the minimum one obtains
\begin{eqnarray}
	\frac{\partial V}{\partial r}&=&-2Al^2r\exp(-l^2r^2)+r=0\nonumber\\
	\Rightarrow r&=&+\phantom{i}l^{-1}\left[\ln(2Al^2)\right]^{1/2},
\end{eqnarray}
where the positive root is chosen and $2Al^2>1$, both to ensure $r>0$. Thus the value of $A$ and $l$ effectively shift the position of the minimum of the potential and hence shift the position of the maximum of the density of the condensate. The effect will be to alter the range of angular velocities for which a particular solitary wave solution exists, however it is expected that $A$ and $l$ will have only a minor effect on the width of the condensate, which is dominated by the harmonic term in (\ref{trap_toroidal}).

\begin{figure}[ht]
\centering
\includegraphics[scale=0.3]{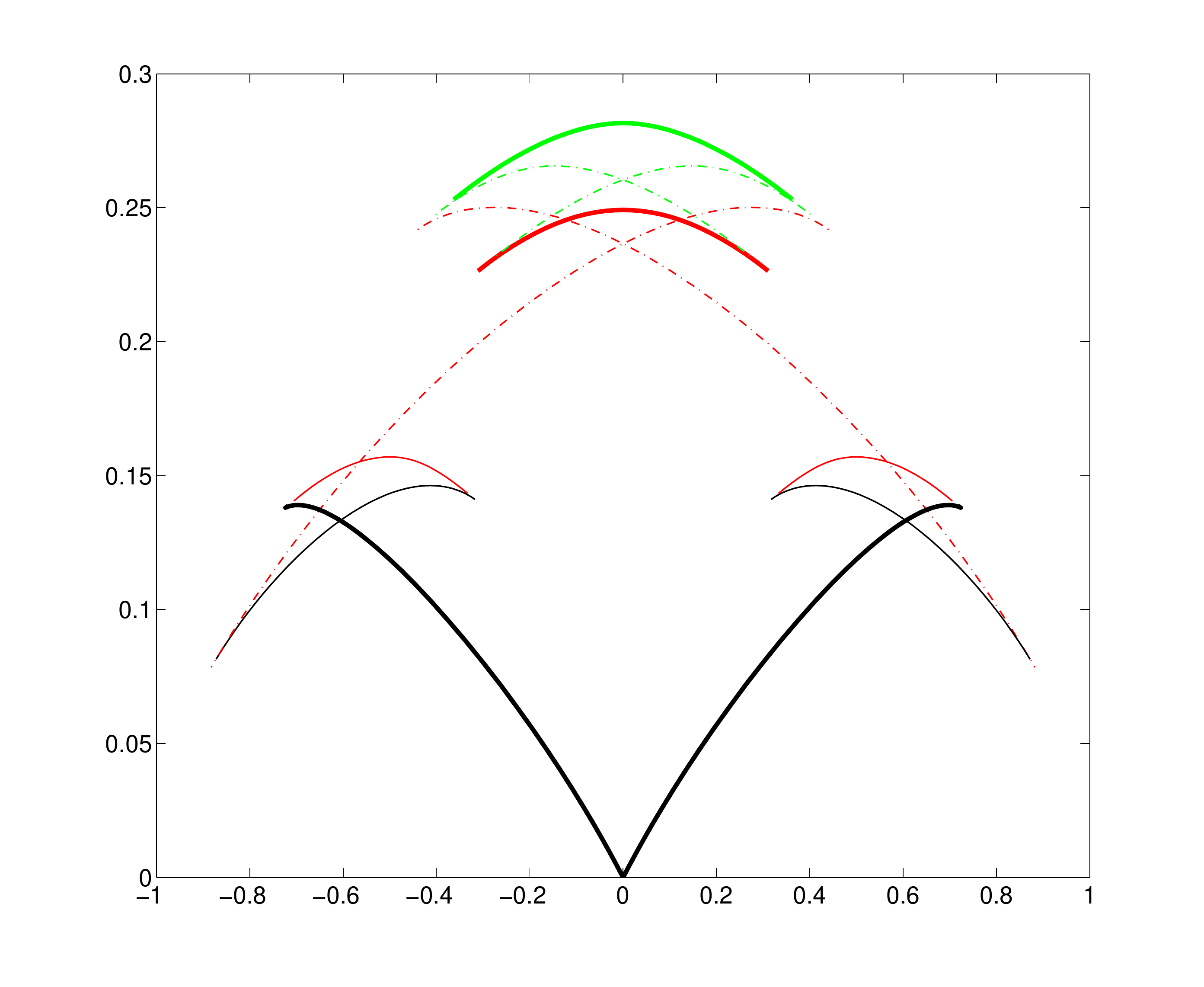}\\
\includegraphics[scale=0.3]{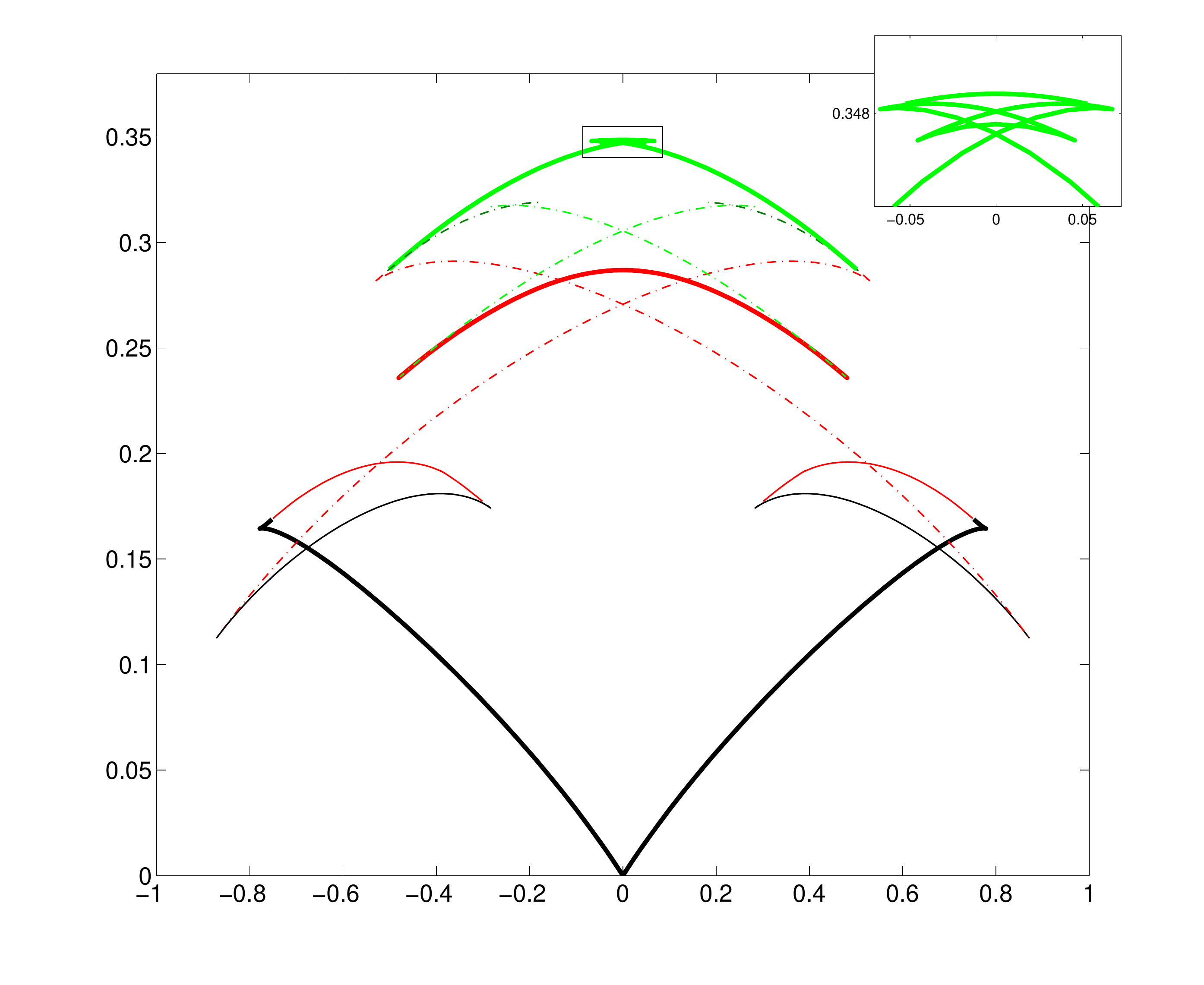}\\
\begin{picture}(0,0)(10,10)
\put(-76,395) {{\tiny{(a)}}}
\put(-76,197) {{\tiny{(b)}}}
\put(-100,125) {$E$}
\put(11,30) {$p$}
\put(-100,322) {$E$}
\put(11,230) {$p$}
\end{picture}
\caption{\baselineskip=10pt \footnotesize (Color online) A comparison of the energy-angular momentum dispersion curve for two different parameter sets with $g=500$. Frame (a) has $\{A,l\}=\{50,0.9\}$ and frame (b) has $\{A,l\}=\{50,1.5\}$. The colors are as in Fig. \ref{tor_500_ep}.}
\label{tor_500_comp_ep}
\end{figure}
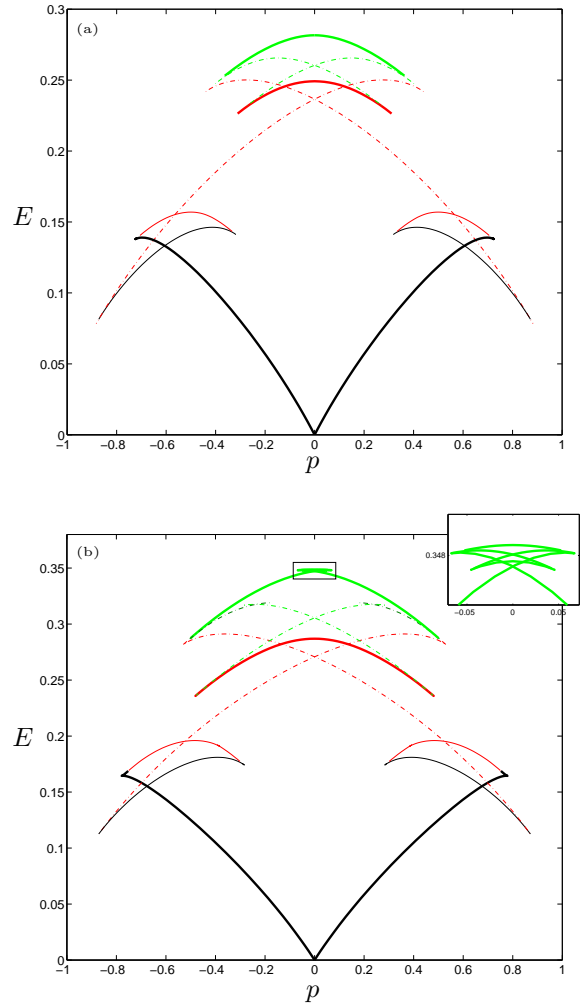

A comparison of the energy-angular momentum dispersion curve for two different parameter sets is shown in Fig.\ \ref{tor_500_comp_ep} for $g=500$ and parameter sets: (a) $\{A,l\}=\{50,0.9\}$ and (b) $\{A,l\}=\{50,1.5\}$. The dispersion curves can then be compared to Fig.\ \ref{tor_500_ep} where $\{A,l\}=\{100,0.9\}$. The value of $g$ for all parameter sets is the same. Parameter sets (a) and (b) vary respectively one of the parameters $\{A,l\}$ of Fig.\ \ref{tor_500_ep} while holding the other fixed. Thus a direct comparison into how $A$ and $l$ effect the dynamics of the condensate can be achieved.

As can be seen in Fig.\ \ref{tor_500_comp_ep} the effect of changing $A$ is almost negligible (compare Fig.\ \ref{tor_500_ep} and Fig.\ \ref{tor_500_comp_ep}(a)). Perhaps the only noticeable difference is a slight increase in the maximum energy, $E_{max}$, as $A$ is decreased. The increase in $E_{max}$ is to be expected since decreasing $A$ will increase the width of the condensate slightly. The effect of changing $l$ is again fairly minimal (compare frames (a) and (b) of Fig.\ \ref{tor_500_comp_ep}). However there are a couple of points that can be made. Firstly, increasing $l$ has the effect of increasing $E_{max}$ fairly considerably (from $E_{max}=0.28$ on frame (a) to $E_{max}=0.35$ on frame (b)). Secondly new features have developed in the dispersion curve for the three (thin dashed-dot light gray (green) curves) and four (thick solid light gray (green) curves) vortex solutions. Ignoring mirror symmetry about $p=0$, the dispersion curve in frame (a) contains a single three vortex configuration and a single four vortex configuration. In contrast the dispersion curve in frame (b) now contains two distinct three vortex configurations and four distinct four vortex configurations (see the blow-up of the region near $E_{max}$ on Fig.\ \ref{tor_500_comp_ep}(b)). Again the new configurations are to be expected; changing $l$ will shift the position of the maximum density of the condensate without altering the width of the condensate much.

The additional configurations that occur by changing $l$ are important, but the essential dynamics of the problem are unaltered; merely a few new configurations to already existing solutions are found (for instance the effect of increasing $g$ might be to create new solutions where greater than four vortices in the condensate are present). Indeed as $g$ is raised to high interaction strengths many more distinct solutions will be present.

\subsection*{D. Discussion}
\label{toroidal_discussion}

\begin{figure}[ht]
\centering
\includegraphics[scale=0.3]{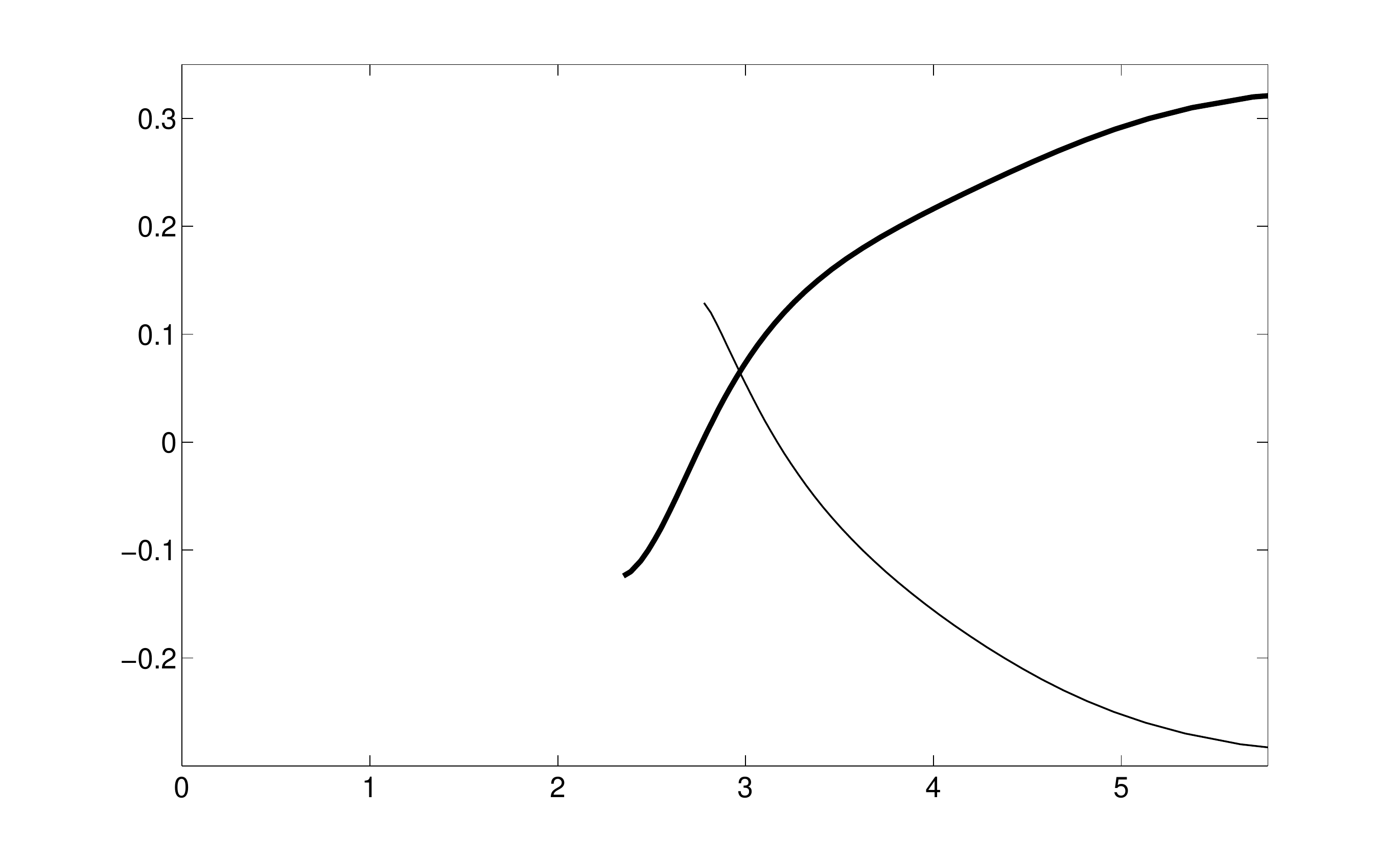}\\
\begin{picture}(0,0)(10,10)
\put(12,22) {$r_0$}
\put(-100,93) {$\Omega$}
\end{picture}
\caption{\baselineskip=10pt \footnotesize The angular velocity $\Omega$ plotted against the distance $r_0$ of the solutions of branches (i) (thick line) and (iii) (thin line) depicted in Fig.\ \ref{tor_500_ep} for $p>0$. The inner edge of the condensate is at $r=R_1=1.22$ and the outer edge of the condensate is at $r=R_2=5.78$.}
\label{branchcomp}
\end{figure}
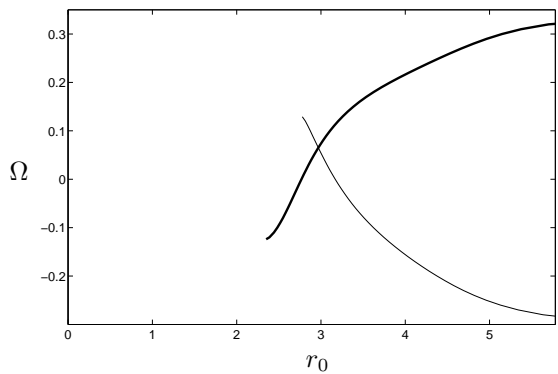

The dynamics of a condensate held under a toroidal trapping potential have been described for two particular cases: $\{g,A,l\}=\{150,40,1\}$ and $\{g,A,l\}=\{500,100,0.9\}$. For $\{g,A,l\}=\{500,100,0.9\}$ the solutions on branch (i) and branch (iii) have both been labelled vortex solutions (see Fig.\ \ref{tor_500_ep}) with the vortex on branch (i) possessing positive circulation and the vortex on branch (iii) possessing negative circulation. The contour plots of the two branches (see Fig.\  \ref{tor_350_cont1} (i) and (iii) respectively) indicate that the solutions are vortices. However this raises an interesting anomaly. Figure \ref{branchcomp} shows the angular velocity against distance plot for branches (i) and (iii). There is one point on the figure in which the curves intersect. At this intersection, the values of the angular velocity and the distances from the center of the condensate are equal for both branches, however the vortices possess different circulations. Such a scenario is mildly paradoxical and is certainly contrary to traditional vortex dynamical ideas. 

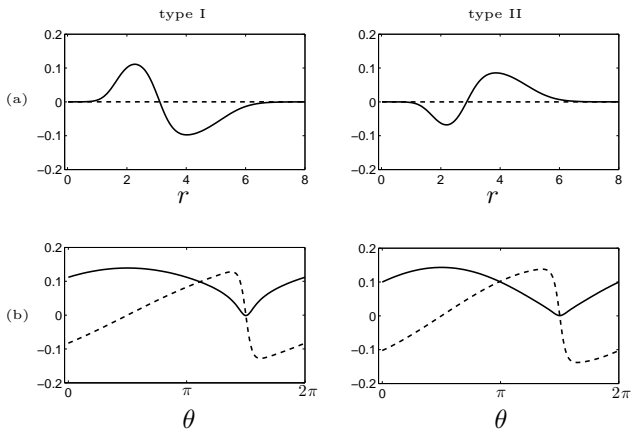
\begin{figure}[ht]
\centering
\includegraphics[scale=0.3]{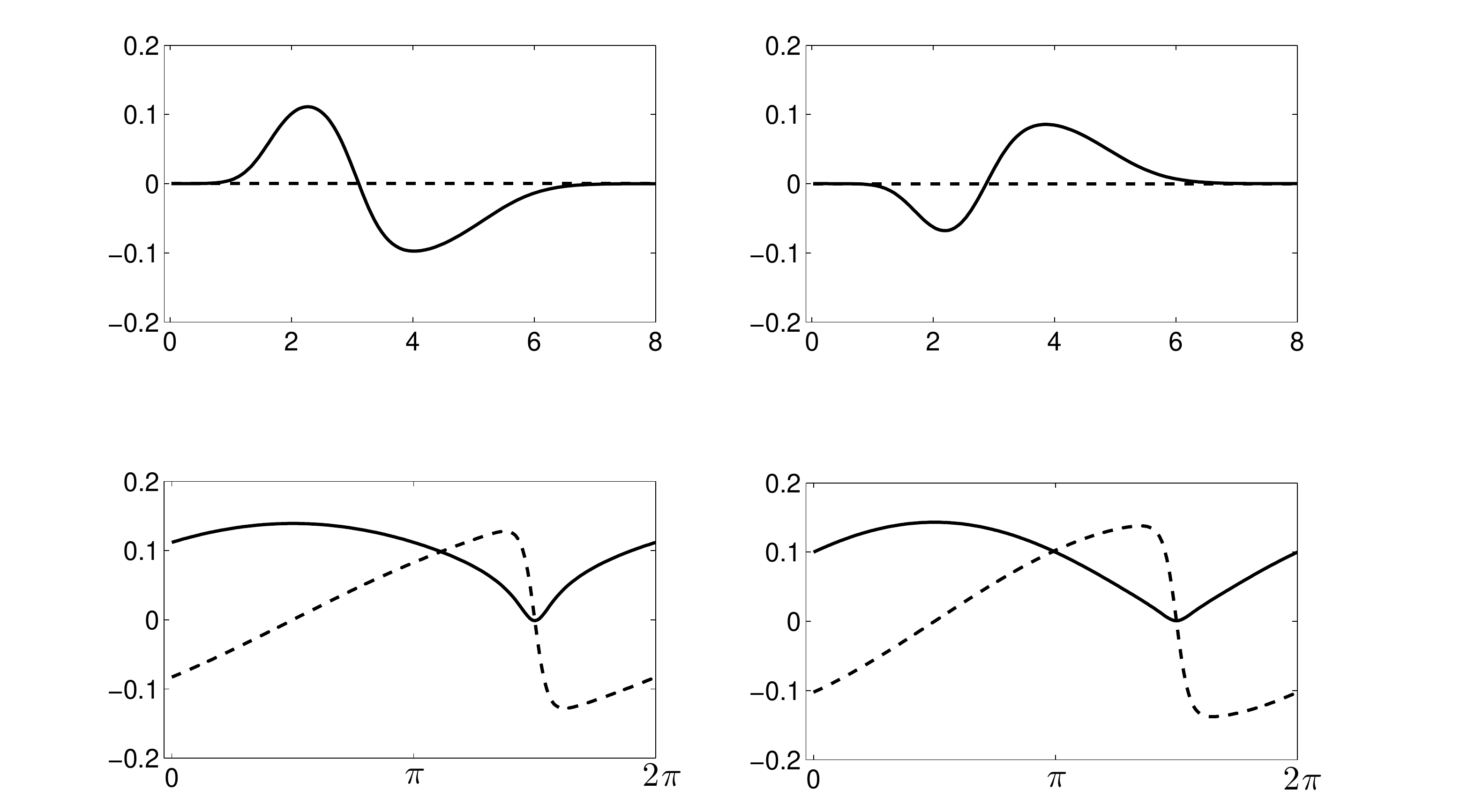}\\
\begin{picture}(0,0)(10,10)
\put(-105,138) {{\tiny{(a)}}}
\put(-105,56) {{\tiny{(b)}}}
\put(-47,170) {{\tiny{type I}}}
\put(70,170) {{\tiny{type II}}}
\put(77,100) {$r$}
\put(-40,100) {$r$}
\put(80,15) {$\theta$}
\put(-38,15) {$\theta$}
\end{picture}
\caption{\baselineskip=10pt \footnotesize Plots of the real (solid line) and imaginary (dashed line) components of the wave function (a) across the direction of propagation and (b) along the direction of propagation of the vortex for $\{g,A,l\}=\{500,100,0.9\}$ and for the two single vortex cases type I and type II (branches (i) and (iii) in Fig.\ \ref{tor_500_ep}, respectively) for $\Omega=0.1$.}
\label{tor_uv}
\end{figure}

A plot of the real and imaginary components of the wave function
($\psi=u+{\rm{i}}v$), Fig.\ \ref{tor_uv} for the two branches shows
that they do indeed contain vortex solutions (with the plots of the
wave function across the direction of propagation confirming that the
vortices have opposite circulation). However, closer inspection of the
plots around $r=7$ shows that the solution of branch (i) depletes to
the ground state at a larger value of $r$ than the solution of branch
(iii). This indicates that branch (iii) possibly contains both a
vortex solution and another embedded solution. Recall that branch
(iii) was created when one of the vortices of branch (ii) reached the
inner edge of the condensate and decayed into sound waves (in fact the
vortex that decays into sound waves is the vortex of branch
(i)). Therefore, the further solitary wave solution contained in branch (iii) is the remnants of the rarefaction waves created when this vortex decayed. 

Thus the solution of branch (i) can indeed be called a (true)
vortex. However the solution of branch (iii) is actually a vortex with
rarefaction waves embedded in the condensate. The statement resolves the apparent paradox discussed earlier. To highlight their differences but at the same time to stress their similarities, the vortex of branch (i) will be denoted `type I' and the vortex and rarefaction waves of branch (iii) will be denoted `type II'. Type II solutions will often be referred to as vortex solutions in order to emphasize the similarities with type I solutions. 

For an annular shaped condensate the method of images will produce two infinite sequences of image vortices. To see this consider a single vortex of positive circulation in the condensate at position $r=r_0$ with the inner edge of the condensate at $R_1$ and the outer edge at $R_2$ so that $R_1<r_0<R_2$. Firstly consider the image of the vortex with the inner edge of the condensate. By the Milne-Thomson circle theorem \cite{newton} the image vortex, named $I_{i1}$, will be of negative circulation and will be at $r=R_1^2/r_0$. The image vortex $I_{i1}$ will itself have an image vortex, of positive circulation, associated with the outer edge of the condensate. The new image will be at $r=R_2^2r_0/R_1^2$ and is named $I_{o2}$. The process can be repeated with $I_{02}$ now producing an image vortex with the inner edge of the condensate and so on. 

Secondly consider the image of the vortex with the outer edge of the condensate. The image vortex, named $I_{o1}$, will be of negative circulation and be at position $r=R_2^2/r_0$. The image vortex $I_{o1}$ will have an associated image vortex with the inner edge of the condensate to produce a vortex of positive circulation, named $I_{i2}$, at $r=R_1^2r_0/R_2^2$. Again the process can be repeated.

The upshot is that two infinite sequences of image vortices are produced. For $r<R_1$, the first two terms of the sequence of image vortices are
\begin{equation}
	\cdots<\frac{R_1^2r_0}{R_2^2}^{(+)}<\frac{R_1^2}{r_0}^{(-)}<R_1,
\end{equation}
where a sign ($\pm$) has been added to indicate the circulation of the vortex. For $R_2<r$, the first two terms of the sequence of image vortices are
\begin{equation}
	\cdots>\frac{R_2^2r_0}{R_1^2}^{(+)}>\frac{R_2^2}{r_0}^{(-)}>R_2.
\end{equation}

It was seen in \cite{mbf} that the effect of a boundary in an inhomogeneous condensate was to alter the position of the image vortex by an amount equal to the depletion in density. The result was used in \cite{mb} for a channel condensate to find an estimate for the velocity of the vortex as a function of distance from the centre of a channel. For the channel condensate there were again two infinite sequences of image vortices. However two distinct vortex solutions were not observed to occur. In the annular shaped condensate there is an asymmetry of the image vortices. The fact that the inner edge of the condensate is depleted by an exponential term and the outer edge of the condensate is depleted by a quadratic term distorts the positions of the image vortices and results in there being two distinct single vortex solutions. Of course there are other solutions that exhibit different configurations, for example $\{g,A,l\}=\{500,50,1.5\}$ detailed in Sect.\ IIIC has four distinct four vortex solutions (see Fig. \ref{tor_500_comp_ep}(b)).

The geometry of the problem unfortunately does not lend itself to easy
calculation of an analytical expression for the angular velocity of a
vortex as a function of distance $r$ from the centre of the
annulus. Various simplifications can be made in the hope of attaining
an expression, among them being to consider only the two closest image
vortices ($I_{i1}$ and $I_{o1}$). However it has been out of reach to
find an analytical expression for the angular velocity.

\section*{IV. STABILITY}

The stability of the solitary wave sequences raises an intriguing question. Because quantized vorticity is an important feature of condensate systems, the manipulation and observation of vortices is a crucial feature of any experiment involving vorticity. Here, a numerical approach is taken to investigate the stability of the solitary wave sequences in Sect.\ III.

Families of solitary wave sequences were found in Sect.\ III for different values of the parameters $\{g,A,l\}$. The stability analysis in this section will concentrate on the parameter set $\{g,A,l\}=\{500,100,0.9\}$ which was described in detail in Sect.\ IIIB. This parameter set is a typical example of the dynamics found in an annular condensate and the energy-angular momentum dispersion curve is displayed in Fig.\ \ref{tor_500_ep}. 

To elucidate the stability of the seven distinct
branches, a time evolution of the found solutions is
performed. Specifically, the time-dependent non-dimensional
Gross-Pitaevskii equation describing the dynamics in the annular
condensate (\ref{gp_tor3}) with toroidal trap (\ref{trap_toroidal}) is
solved using a 4th order finite difference and 4th order Runge-Kutta
scheme. The stationary solutions found in Sect.\ IIIB were
found on an irregular polar grid. To translate the
solutions to a regular cartesian grid, cubic splines are used. The
solutions in each branch, for various angular velocities, are evolved
in time until $t\sim150$, a long enough time period to ensure a number
of full revolutions of the solutions. It is expected that, over the
course of a few revolutions, the unstable solutions will show one of
two features: either they will decay into sound waves or, in the case
of vortices will drift towards the inner or outer edge of the
condensate. The stable solutions are expected to be unchanged after
the time evolution.  It is emphasised that the approach taken here
does not prove the stability, or instability, of the respective
branches of the energy-angular momentum dispersion curve, but this approach gives a strong indication to the respective stability of the different solutions.

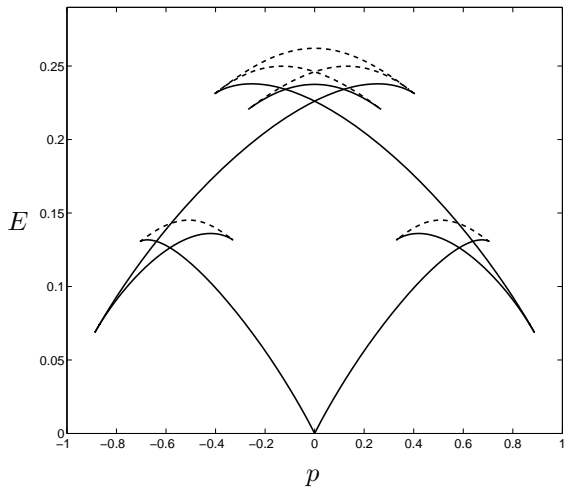
\begin{figure}[ht]
\centering
\includegraphics[scale=0.3]{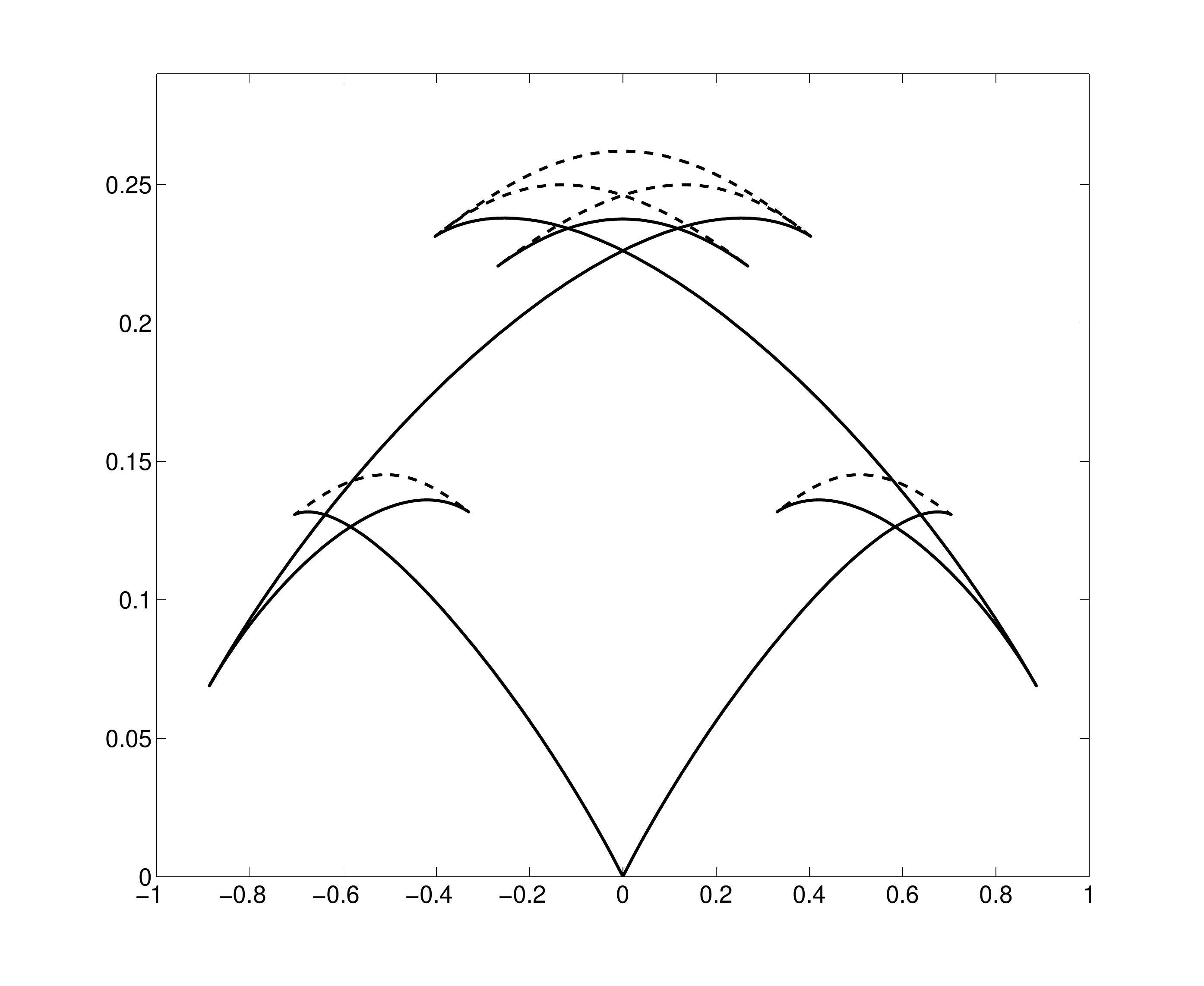}\\
\begin{picture}(0,0)(10,10)
\put(11,26) {$p$}
\put(-102,121) {$E$}
\end{picture}
\caption{\baselineskip=10pt \footnotesize The energy-angular momentum dispersion curve for the annular condensate for parameter set $\{g, A, l\}=\{500, 100, 0.9\}$ of Fig.\ \ref{tor_500_ep}. The relative stability of the different branches are indicated; stable (solid line), unstable (dashed line).}
\label{stab}
\end{figure}

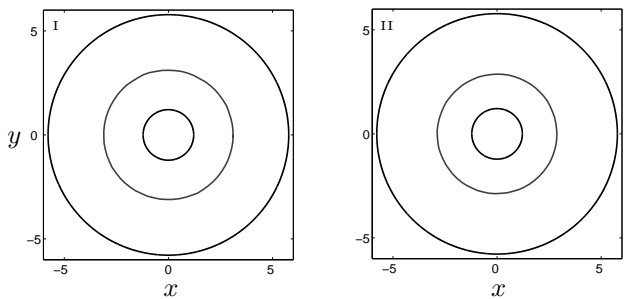
\begin{figure}[ht]
\centering
\includegraphics[scale=0.3]{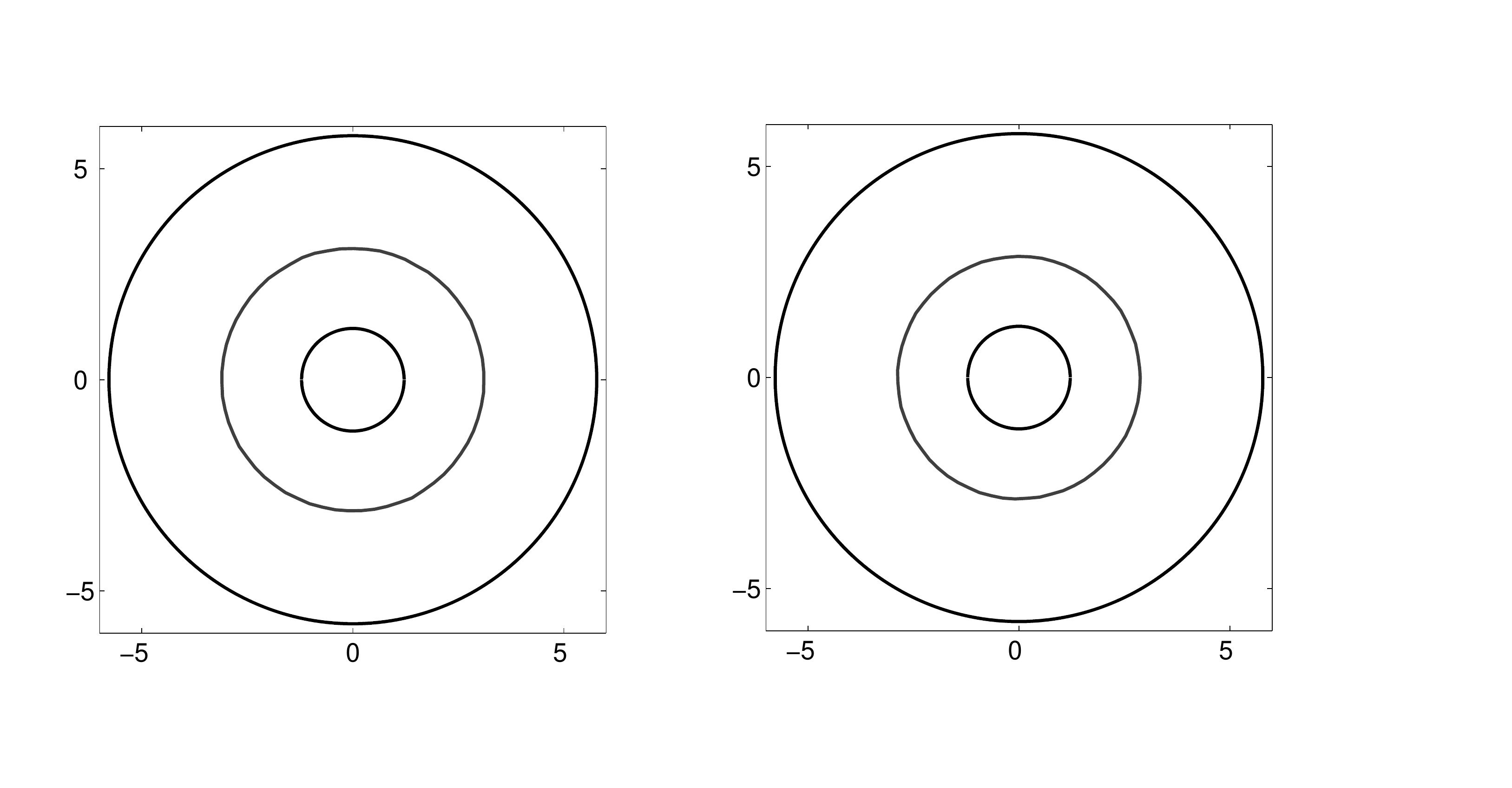}\\
\begin{picture}(0,0)(10,10)
\put(-108,95) {$y$}
\put(-49,37) {$x$}
\put(75,37) {$x$}
\put(-91,138) {{\tiny{I}}}
\put(33,138) {{\tiny{II}}}
\end{picture}
\caption{\baselineskip=10pt \footnotesize The path (gray lines) of the single vortex solutions of type I and type II for $\Omega=0.1$ for integration forward in time. The vortices are at $r=3.11$ and $r=2.87$ respectively and are expected to be stable. The edges of the condensate are the black lines and are at $r=R_1=1.22$ and $r=R_2=5.78$.}
\label{stab_single}
\end{figure}

Of the seven branches of Fig.\ \ref{tor_500_ep}, it is observed that four branches are stable; branches (i), (iii), (iv) and (vii), while three branches are unstable; branches (ii), (v) and (vi). The relative stability of the different branches is summarised in Fig.\ \ref{stab}. The motion of the single vortex of type I and II (branches (i) and (iii) of Fig.\ \ref{tor_500_ep}, respectively) are both stable. A plot of the position of the vortex in each case is shown in Fig.\ \ref{stab_single} for angular velocity $\Omega=0.1$. One of the other stable branches, branch (iv), will be considered in more detail in Sect.\ VA. 

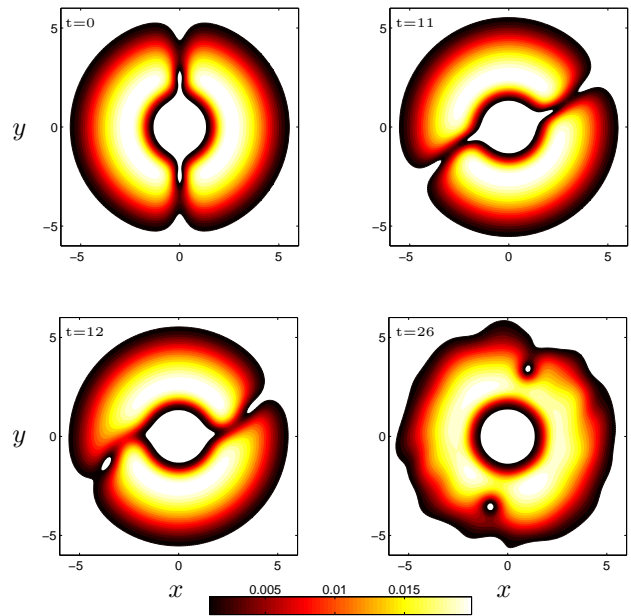
\begin{figure}[ht]
\centering
\includegraphics[scale=0.4]{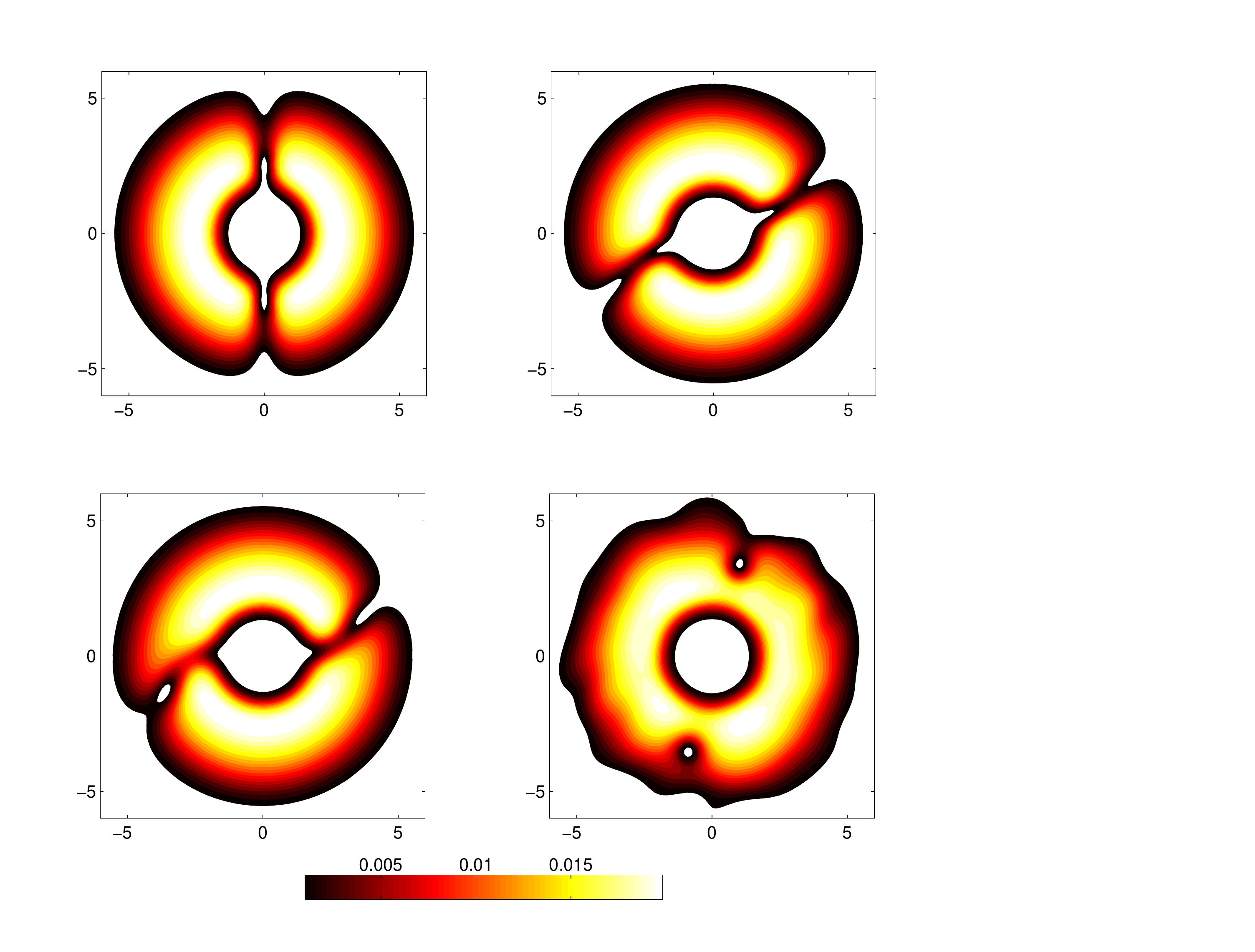}\\
\begin{picture}(0,0)(10,10)
\put(-83,259) {{\tiny{t=0}}}
\put(42,259) {{\tiny{t=11}}}
\put(-83,142) {{\tiny{t=12}}}
\put(42,142) {{\tiny{t=26}}}
\put(-103,218) {$y$}
\put(-103,102) {$y$}
\put(-44,43) {$x$}
\put(80,43) {$x$}
\end{picture}
\caption{\baselineskip=10pt \footnotesize (Color online) Contour profiles of the stationary solution for $\Omega=0.1$ of branch (vi) for four vortices placed along $\theta=\pm\pi/2$ at different intervals in time. The decay of the solution over time indicates instability.}
\label{stab_4}
\end{figure}

Turning now to the unstable solutions, branch (vi), where there are four vortices of alternate circulation along $\theta=\pm\pi/2$, is considered as an example of the dynamics observed. Figure \ref{stab_4} shows snapshots of the contour profile at different intervals of time for angular velocity $\Omega=0.1$. As is seen, slight disturbances to the velocity profile cause the four vortex solution to decay. Initially, instabilities occur ($t=11$), causing the four vortices to decay into two single vortices and sound waves ($t=12$). After a short time interval, two vortices are observed to remain in the condensate which is now filled with sound waves ($t=26$). The decay of the vortices in branches (ii) and (v) is similar.

The relative stabilities of the seven branches is to be expected. Vortex rings/pairs in an infinite homogeneous condensate were found to be stable \cite{jpr1986,b2004} as well as vortex rings/pairs in a semi-infinite channel condensate \cite{kp2}. However multiple vortex rings/pairs are unstable in these geometries. 
For the case considered in this paper, the unstable solutions are ones for which there exists two (i.e. multiple) vortices along the same polar angle. The two vortices will decay into a single vortex and sound waves (see Fig.\ \ref{stab_4}). In contrast, solutions that contain only a single vortex along a polar angle are stable. Indeed, for branch (vi) where there are two vortices along $\theta=\pi/2$ and one vortex along $\theta=-\pi/2$, the instability develops along $\theta=\pi/2$ while the single vortex along $\theta=-\pi/2$ remains stable. However, over time the sound waves from the decay of the two vortices along $\theta=\pi/2$ propagate through the condensate and render the single vortex unstable. 

\section*{V. EVOLUTION AND COLLISIONS OF SOLITARY WAVES}

The evolution of vortices in multiply connected domains has received much attention over the years. The majority of the focus has been on modelling classical fluid dynamical situations of interest, for example the motion of a vortex around islands \cite{c1} or through a gap in a wall \cite{c2,rj}. Traditional hydrodynamical methods for solving such situations revolve around a point vortex model which enables a reduction in the dimension of the system whilst capturing all the essential dynamics. Point vortex models can also be successfully applied to annular condensates; see \cite{zueva,lakaniemi,sssl,mssss}. However the point vortex model relies on incompressible dynamics so the density in the bulk of the condensate is constant and is zero outside of the condensate. Thus, interactions between the background condensate and the vortices are forbidden and so the transfer of energy between the vortices and the background condensate does not occur. Instead, vortices initially in the condensate will remain in the condensate. Decay of vortices is therefore not possible using a `boxlike' trap and a point vortex model. It has already been seen (Sect.\ IV) that interactions with the background condensate can have a significant effect on the dynamics. Figure \ref{stab_4} shows that the interactions between the background condensate and the four vortices results in an instability developing and the decay of two of the vortices into sound waves. 

Experimentally, collisions of three-dimensional solitary waves have been observed in \cite{ginsberg} and a recent theoretical study \cite{kombrand} considered the head on collisions of vortex rings and solitons in a cigar shaped condensate. It was observed in \cite{kombrand} that, depending on the speed of approach, the interactions of the solitary waves can be elastic or inelastic, with the solitary waves either repelling, passing through or annihilating each other. Collisions where the solitary waves pass through each other (elastic collision) is also forbidden by point vortex models. 

In a recent paper, Li et al. \cite{lhk} considered vortex-antivortex pairs in a two-dimensional harmonically trapped condensate. In such a condensate it becomes difficult to simulate vortex-vortex or vortex-antivortex collisions. Instead a toroidal geometry explicitly allows this possibility. This section will consider the evolution of vortex-vortex pairs and vortex-antivortex pairs in the annular condensate by numerically simulating the GP equation (\ref{gp_tor3}) by the same method employed in Sect.\ IV. It is noted that, by solving Eq.\ (\ref{gp_tor3}) with toroidal trap (\ref{trap_toroidal}), the bulk condensate density is not constant and thus the simulations in this section will provide a comparison with the point vortex models employed in previous papers \cite{zueva,lakaniemi,sssl,mssss}. 

The parameters of the condensate are again chosen to be $\{g,A,l\}=\{500,100,0.9\}$. The evolution and collisional dynamics of two vortices are the main focus and thus the initial states for the condensate will contain two vortices, one along $\theta=\pi/2$ and one along $\theta=-\pi/2$. As detailed before, there exist two distinct single vortex solutions denoted type I and type II (branch (i) and branch (iii) on Fig.\ \ref{tor_500_ep}, respectively). The solutions on these branches are isolated (note that it is not the vortex solution that is isolated, but the complete solution in the absence of the ground state). The initial state can then be formed from $\psi=\psi_0\psi_{v_+}\psi_{v_-}$, for ground state $\psi_0$ and vortex states $\psi_{v_{\pm}}$ (\ref{ansatz}) for vortices along $\theta=\pm\pi/2$ respectively. Different combinations and different circulations of the single vortex solutions of type I and type II are considered.

\subsection*{A. Vortex-Vortex Pair}

\begin{figure}[ht]
\centering
\includegraphics[scale=0.3]{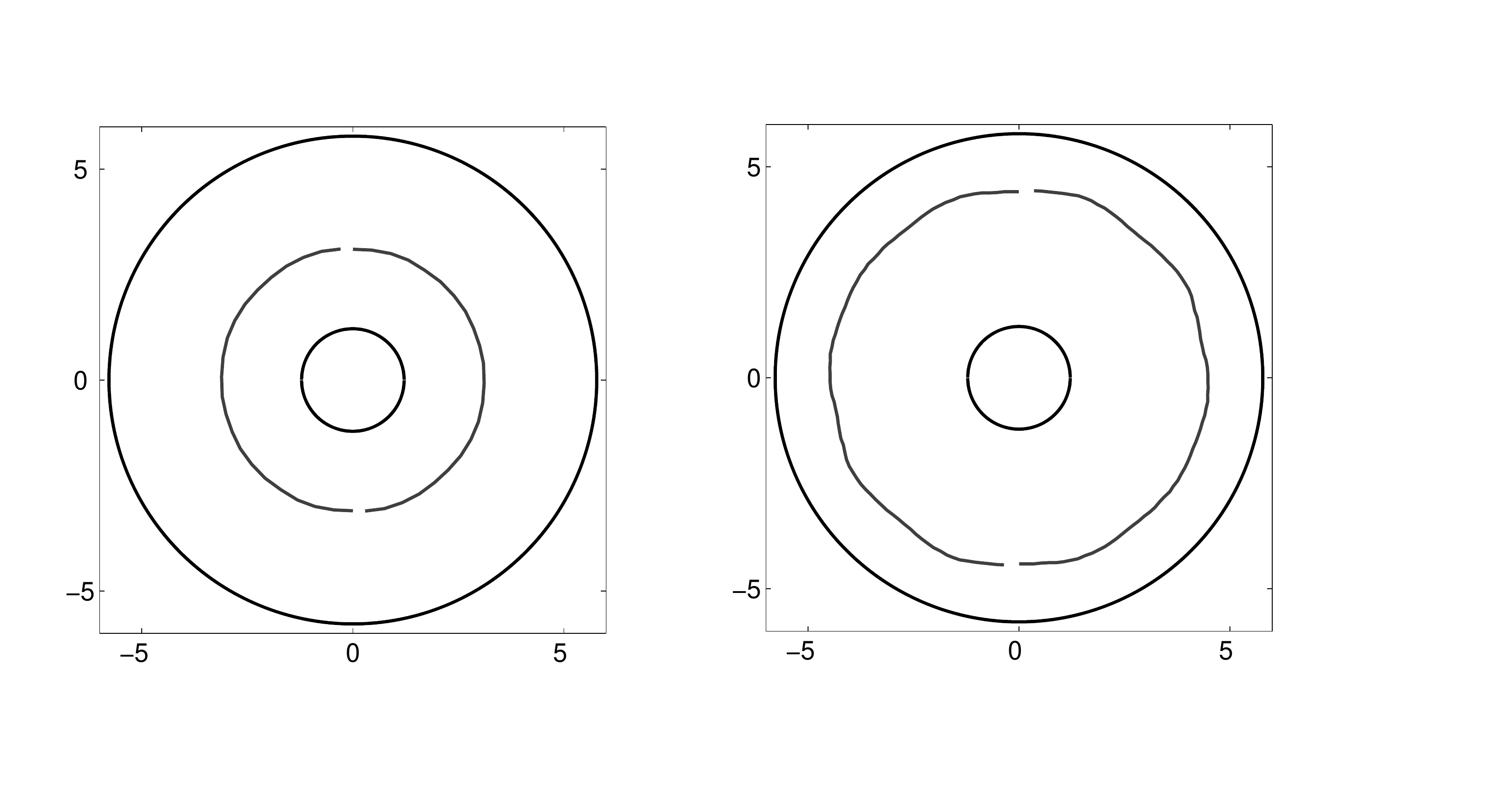}\\
\begin{picture}(0,0)(10,10)
\put(-108,95) {$y$}
\put(-49,37) {$x$}
\put(75,37) {$x$}
\put(-91,138) {{\tiny{(a)}}}
\put(33,138) {{\tiny{(b)}}}
\end{picture}
\caption{\baselineskip=10pt \footnotesize The path (gray lines) of vortices placed initially on opposite sides of the condensate with the same circulation. Frame (a) contains two vortices of type I of Fig.\ \ref{tor_500_ep} at $r_0=3.82$ at $\theta=\pm\pi/2$ and frame (b) contains two vortices of type II at $r_0=4.38$. The edges of the condensate are the black lines and are at $r=R_1=1.22$ and $r=R_2=5.78$.}
\label{vv_1}
\end{figure}
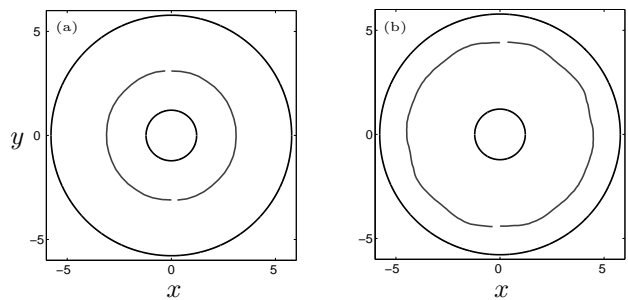

Consider first the evolution of two vortices of type I of the same circulation and placed at equal radii on opposite sides of the condensate. As expected, the vortices traverse paths of constant radius and move with the same angular velocity. The paths of the vortices are shown in Fig.\ \ref{vv_1}(a) for vortices placed at $r_0=3.82$ where the angular velocity is $\Omega=0.2$. Also shown in Fig.\ \ref{vv_1}(b) is the identical situation but now for vortices of type II placed at radii $r_0=4.38$ where the angular velocity is $\Omega=-0.2$. Note that the construction of these models forms condensates that are identical to branch (vi) of the dispersion curve in Fig.\ \ref{tor_500_ep} which was shown to be numerically stable. However, for a given radii of the two vortices, the angular velocity is different in all cases. 

\begin{figure}[ht]
\centering
\includegraphics[scale=0.4]{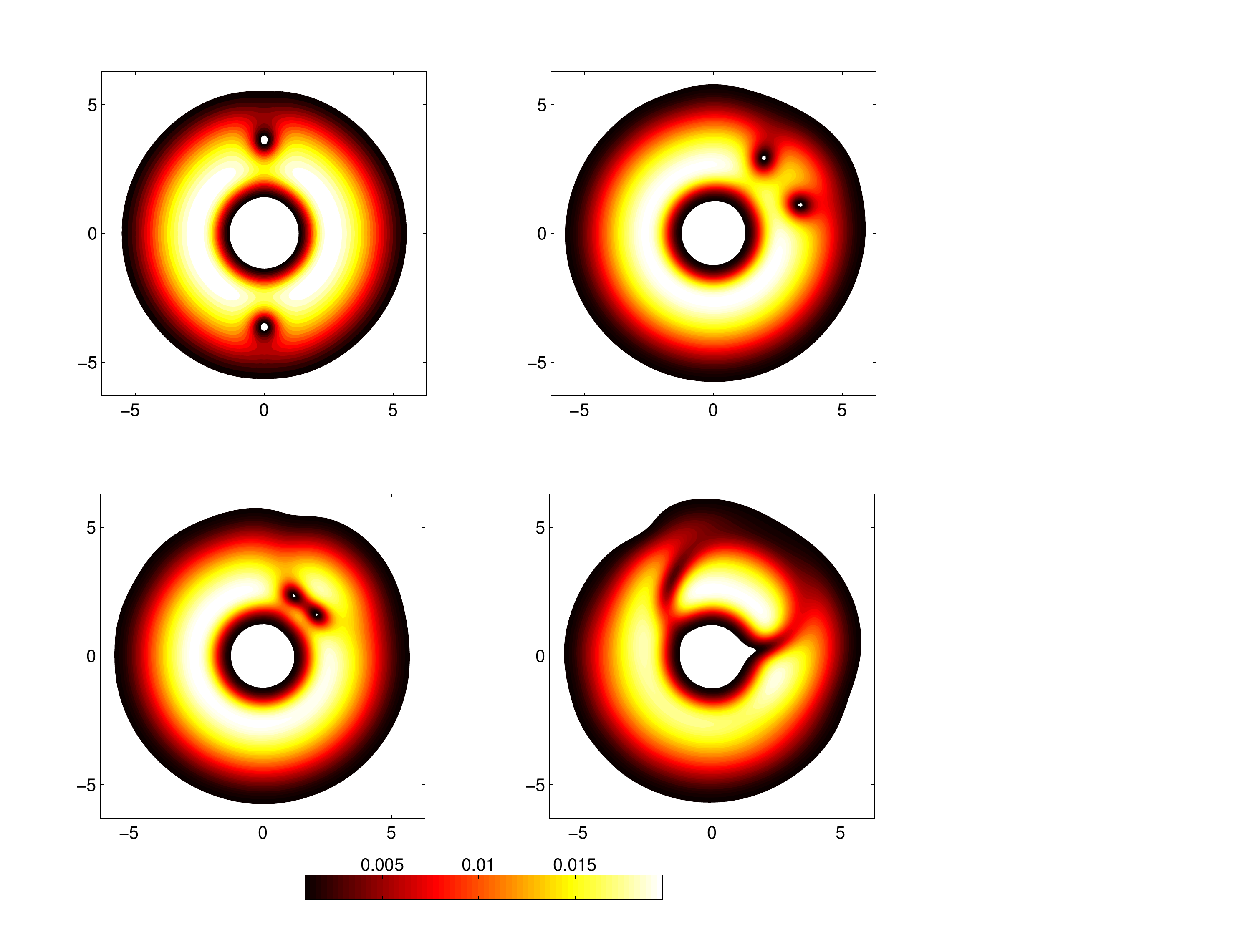}\\
\begin{picture}(0,0)(10,10)
\put(-83,259) {{\tiny{t=0}}}
\put(42,259) {{\tiny{t=8}}}
\put(-83,142) {{\tiny{t=10}}}
\put(42,142) {{\tiny{t=12}}}
\put(-103,218) {$y$}
\put(-103,102) {$y$}
\put(-44,43) {$x$}
\put(80,43) {$x$}
\end{picture}
\caption{\baselineskip=10pt \footnotesize (Color online) Contour profiles of the evolution of one vortex of type I placed at $\theta=-\pi/2$ of Fig.\ \ref{tor_500_ep} and one vortex of type II at $\theta=\pi/2$. Both vortices are initially are at $r=3.62$ and have the same circulation.}
\label{vv_2}
\end{figure}
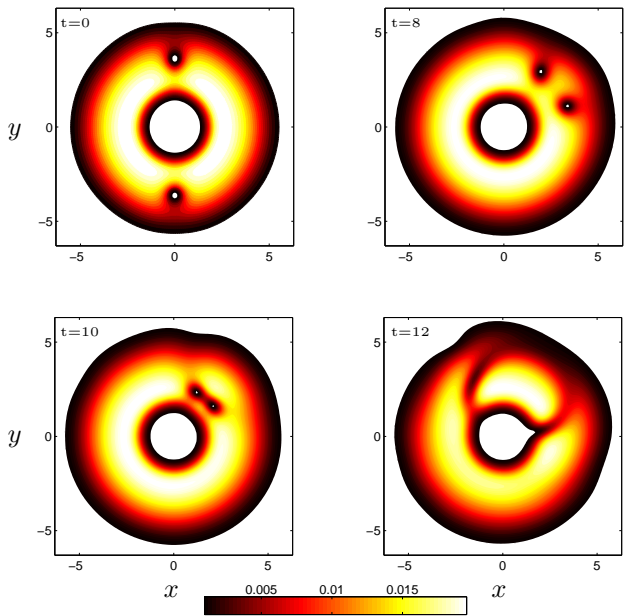

Next, the interactions of type I and type
  II vortices are considered.
A vortex of type I at $\theta=-\pi/2$ and a vortex of type II at $\theta=\pi/2$ are placed at the same radial positon $r_0=3.62$. The vortices have the same circulation, however at $r_0=3.62$, the vortex of type I has positive angular velocity ($\Omega=0.18$), whereas the vortex of type II has negative angular velocity ($\Omega=-0.1)$. Thus the vortices are expected to move towards one another but with different magnitudes of angular velocity. Indeed, as time evolves, the vortices begin to approach each other along the same radii with their initial angular velocities. However as the distance between the vortices decreases, the vortices begin to lose angular velocity and move towards the centre of the condensate. They continue to move towards the centre of the condensate until they reach  the inner boundary where they both decay into sound waves. A selection of snaps of the contour profile in Fig.\ \ref{vv_2} at different time intervals shows the process.

\subsection*{B. Vortex-Antivortex Pair}

\begin{figure}[ht]
\centering
\includegraphics[scale=0.3]{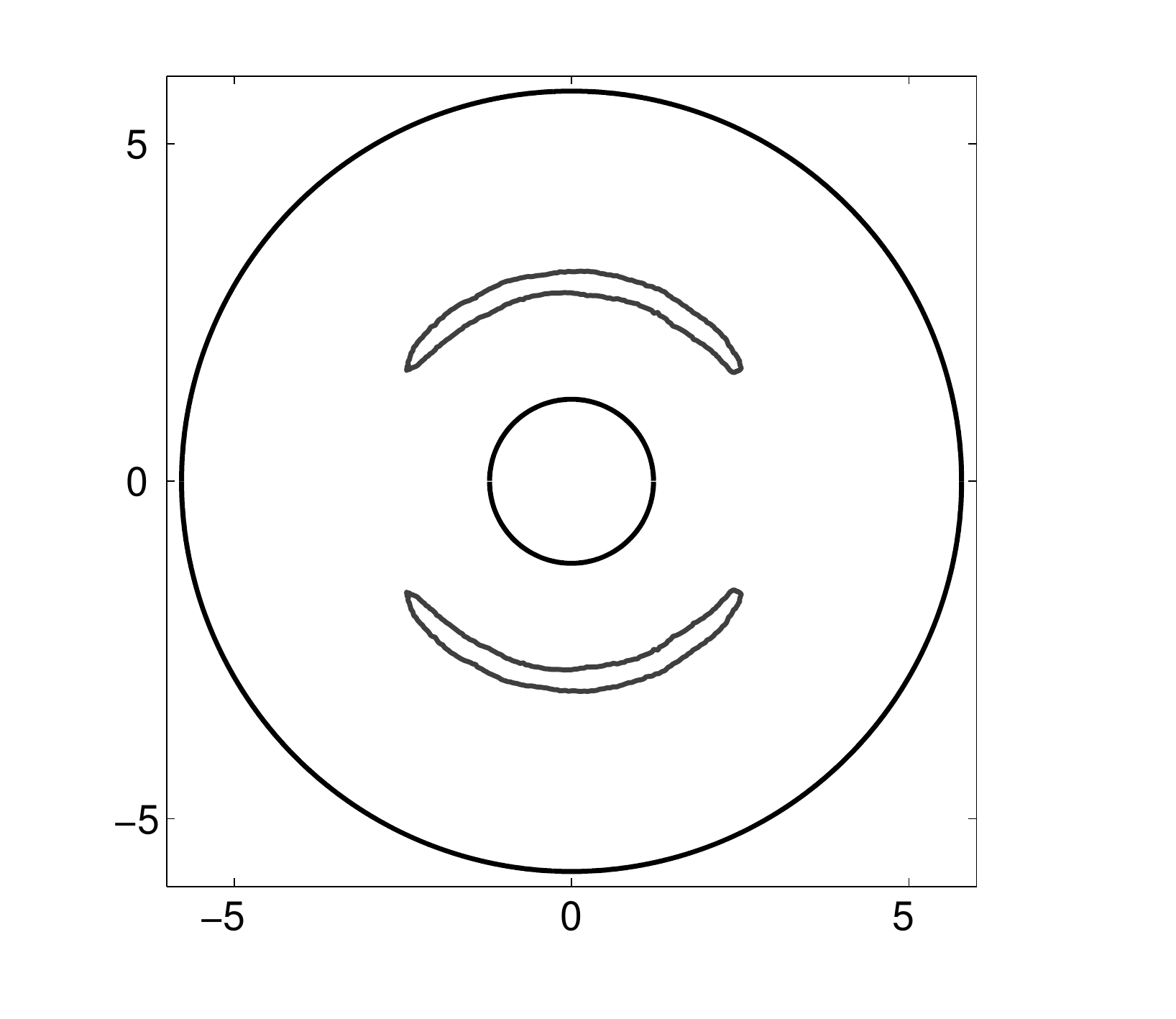}\\
\begin{picture}(0,0)(10,10)
\put(-57,85) {$y$}
\put(5,23) {$x$}
\end{picture}
\caption{\baselineskip=10pt \footnotesize The path (gray lines) of vortices of type I of Fig.\ \ref{tor_500_ep} placed initially on opposite sides of the condensate with opposite circulation and with radius $r_0=3.11$. The edges of the condensate are the black lines and are at $r=R_1=1.22$ and $r=R_2=5.78$. The angular velocity is $\Omega=0.1$.}
\label{va_1}
\end{figure}
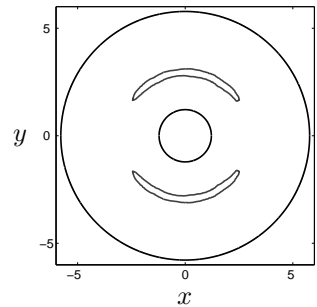

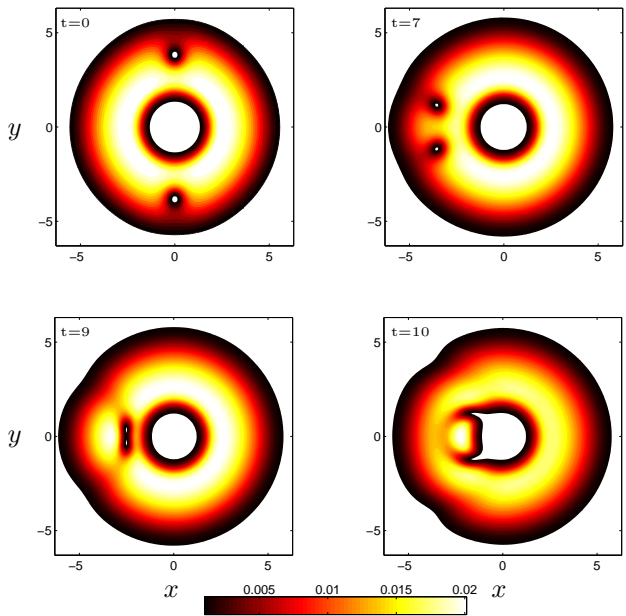
\begin{figure}[ht]
\centering
\includegraphics[scale=0.4]{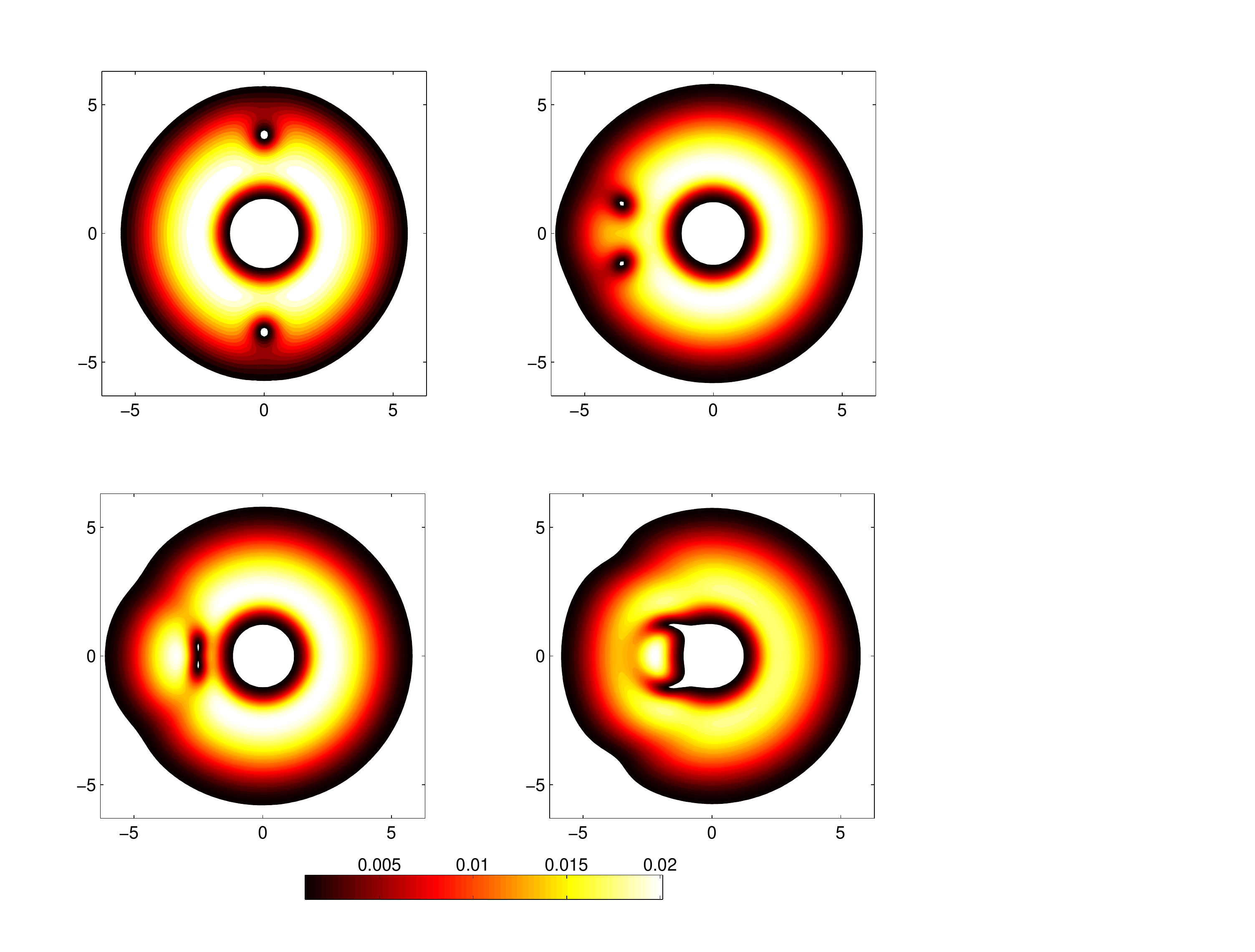}\\
\begin{picture}(0,0)(10,10)
\put(-83,259) {{\tiny{t=0}}}
\put(42,259) {{\tiny{t=7}}}
\put(-83,142) {{\tiny{t=9}}}
\put(42,142) {{\tiny{t=10}}}
\put(-103,218) {$y$}
\put(-103,102) {$y$}
\put(-44,43) {$x$}
\put(80,43) {$x$}
\end{picture}
\caption{\baselineskip=10pt \footnotesize (Color online) Contour profiles of the evolution of vortices of type I of Fig.\ \ref{tor_500_ep}. Both vortices are initially are at $r_0=3.82$ and have opposite circulation. The angular velocity is $\Omega=0.2$.}
\label{va_2}
\end{figure}

\begin{figure}[ht]
\centering
\includegraphics[scale=0.4]{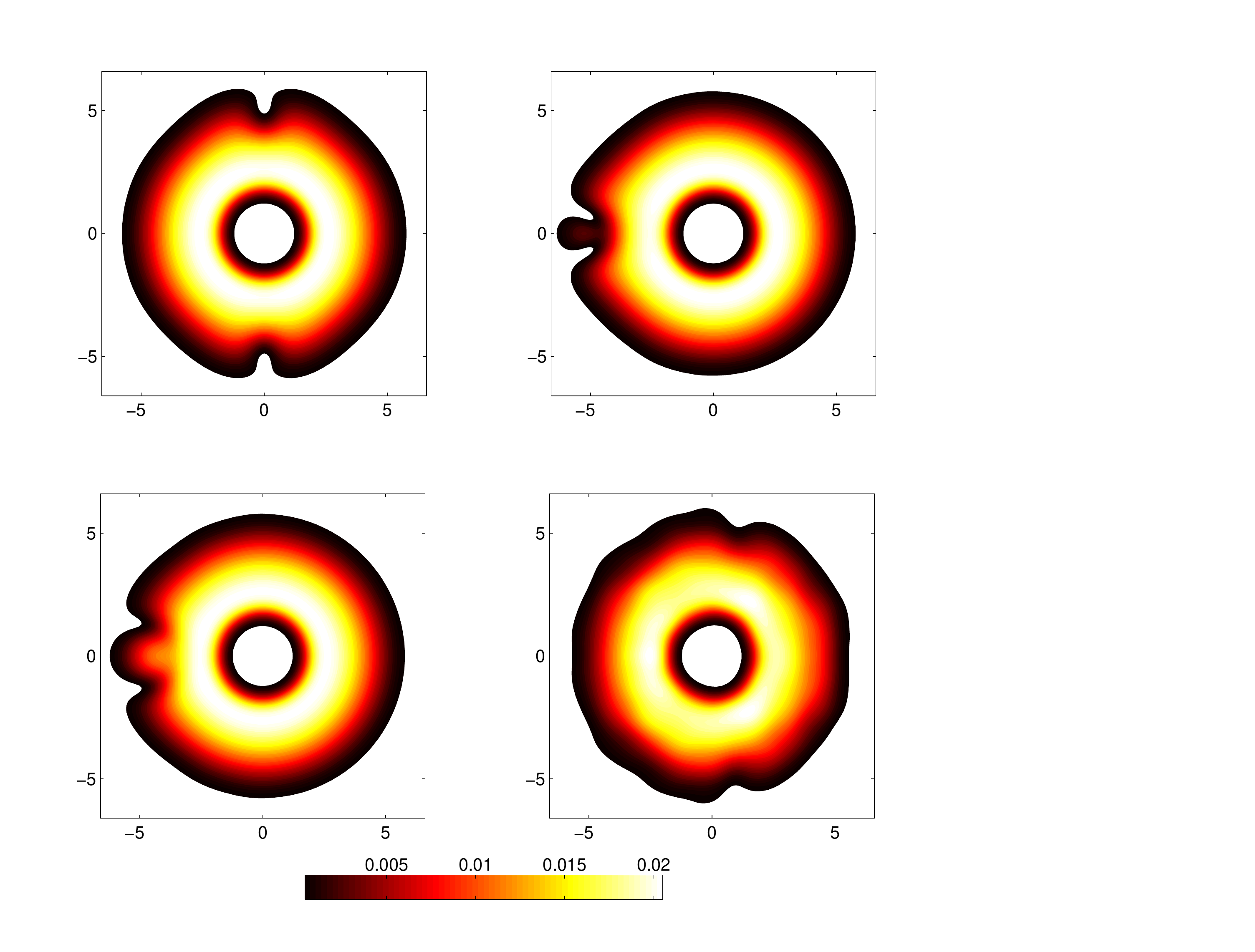}\\
\begin{picture}(0,0)(10,10)
\put(-83,259) {{\tiny{t=0}}}
\put(42,259) {{\tiny{t=5}}}
\put(-83,142) {{\tiny{t=6}}}
\put(42,142) {{\tiny{t=150}}}
\put(-103,218) {$y$}
\put(-103,102) {$y$}
\put(-44,43) {$x$}
\put(80,43) {$x$}
\end{picture}
\caption{\baselineskip=10pt \footnotesize (Color online) Contour profiles of the evolution of vortices of type I of Fig.\ \ref{tor_500_ep}. Both vortices are initially are at $r_0=5.15$ and have opposite circulation. The angular velocity is $\Omega=0.3$. The vortices pass through each other between $t=5$ and $t=6$.}
\label{va_3}
\end{figure}
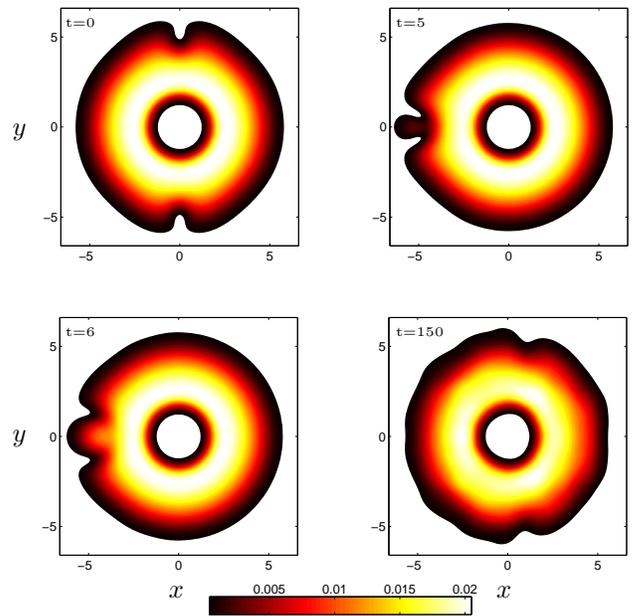

The collision of the two vortices of type I and type II, detailed in the previous section and in Fig.\ \ref{vv_2} has opened up the question as to whether there is an analogy with the collisional properties outlined by Komineas \& Brand \cite{kombrand}. For solitary waves in a cigar shaped trap, the head-on collisions were elastic for low and high velocities of approach but were inelastic for intermediate values of velocity. To investigate the elasticity of the collisions in an annular condensate, a series of simulations for different angular velocities were considered for two vortices both taken to be of type I and placed on opposite sides of the condensate at the same radius but with opposite circulations. The three angular velocities chosen are $\Omega=0.1$ (with corresponding $r_0=3.11$), $\Omega=0.2$ ($r_0=3.82$) and $\Omega=0.3$ ($r_0=5.15$). The three simulations are outlined in Fig.'s \ref{va_1}, \ref{va_2} and \ref{va_3} respectively. 

When the angular velocity is small ($\Omega=0.1$), initially the vortices approach one another with constant angular velocity. However as the distance between the vortices begins to reduce, their influence on each other begins to take effect and dominate the flow in the condensate. At such time, the angular velocity is reduced and the vortices move to a new radius where they begin to separate again at a new, constant, angular velocity. The process is repeated when the vortices next meet at the opposite side of the condensate, however now the vortices move back to the original radius and angular velocity. The process continues forever and is therefore elastic: the vortices upon collision merely transfer to a new energy and angular momentum configuration.  In Fig.\ \ref{va_1}, a plot of the paths of the vortices is provided.

As the angular velocity is increased to $\Omega=0.2$, the vortices again upon collision slow down and move to a smaller radius. However, the vortices now possess too much energy to simply find a new radius to move along. Instead the vortices propel each other towards the inner edge of the condensate where they decay into sound waves. A similar process was observed to occur for the vortex-vortex pair considered in Fig.\ \ref{vv_2}. Contour plots in Fig.\ \ref{va_2} at different time intervals for $\Omega=0.2$ show the inelastic collision and decay process. 

Further increasing the angular velocity will cause the collisions to become elastic again. Taking $\Omega=0.3$ as a typical example of the dynamics observed, the vortices now pass through one another and continue in their original direction of motion. However, during each collision a fraction of the energy of the vortices is lost to sound waves. Over time, therefore, the vortices slowly begin to lose energy. The energy lost during the collision to sound waves can be determined by
\begin{equation}
\Delta E=\frac{E_a-E_b}{E_a}
\end{equation}
where $E_a$ is the energy before collision and $E_b$ is the energy after collision and where $E_b$ is inferred from the angular velocity of the vortices after the collision. The value of the energy loss has been calculated after a number of collisions for $\Omega=0.3$ and is found to be always less than $1\%$. Figure \ref{va_3} shows contour plots of the initial collision (between $t=5$ and $t=6$) and a plot of the state of the system after a long evolution ($t=150$).

The correspondence of the dynamics with the dynamics observed in \cite{kombrand} for the cigar shaped condensate is evident. For the annular condensate, high and low angular velocities possess elastic collisions, whereas intermediate values of angular velocity possess inelastic collisions. 

Another striking feature of the interaction of the dynamics in the annular condensate is the similarity of Fig.'s \ref{stab_single}, \ref{vv_1} and \ref{va_1} to the dynamics observed in \cite{zueva,lakaniemi} by a point vortex model. Here, the figures show the motion of the vortices as time is evolved. In inhomogeneous condensates where the density varies over all space, the motion of the vortices is because of the density gradients' depletion at the boundary (see \cite{mb}). The model detailed in Sect.\ IV and this section to find the evolution of vortices considers the depletion of the density gradient. However, the point vortex model does not take into account the density gradient. The two models give the same trajectories but different velocities for each trajectory for both type I and II solutions. The point vortex model therefore provides a qualitative but not quantitative representation of the dynamics of the elastic collisions. Furthermore the point vortex model will not be able to capture any of the inelastic collisions. 

\section*{VI. PERSISTENT FLOW}

A persistent flow (supercurrent) has been observed to occur in a toroidal trap by \cite{racnhp}, where the creation of vortices was also seen. The dynamics of solitary waves in a toroidal trap with a persistent flow present are readily obtained by a simple adaptation of the theory presented in Sect.\ II. As noted before, a persistent flow can be characterised by the existence of a vortex at the origin. Therefore take the non-dimensional GP equation (\ref{gp_tor3}) with the same trapping potential (\ref{trap_toroidal}). Now consider the transformation of the wave function $\hat{\psi}=\psi\exp({\rm{i}}\theta)$ which is equivalent to placing a singly quantized vortex at the origin. Then, the modified GP equation now reads
\begin{equation}
	\label{gp_tor1_sc}
	2{\rm i}\frac{\partial\hat{\psi}}{\partial t}=-\nabla^2\hat{\psi}+\frac{2{\rm i}}{r^2}\frac{\partial\hat{\psi}}{\partial \theta}-\left(\mu-V(r)-g|\hat{\psi}|^2-\frac{1}{r^2}\right)\hat{\psi}.
\end{equation}
Transfer to the rotating frame using $\theta'=\theta-\Omega t$, so that $\partial/\partial t\rightarrow-\Omega\partial/\partial\theta'$, and such that (\ref{gp_tor1_sc}) becomes
\begin{equation}
	\label{gp_tor_sc}
	2{\rm i}\frac{\partial\psi}{\partial\theta}\left(\Omega+\frac{1}{r^2}\right)=\nabla^2\psi+
	\left(\mu-V(r)-g|\psi|^2-\frac{1}{r^2}\right)\psi,
\end{equation}
where the hat ($\hat{\phantom{e}}$) notation for $\psi$ and the dash ($'$) notation for $\theta$ has been dropped for convenience of notation. From now on $\psi$ will refer to the wave function in the condensate with a persistent flow. Equation (\ref{gp_tor_sc}) describes the dynamics of solitary wave solutions in a toroidal trap with a persistent flow. Solitary wave solutions that move with a constant angular velocity $\Omega$ at constant $r$ are sought by the same numerical procedure introduced in Sect. \ III.

The normalisation condition (\ref{tor_norm}) is unchanged;
\begin{equation}
	\label{tor_norm_sc}
	\int_{\mathcal{V}}|\psi|^2r\,drd\theta=2\pi\int_{\mathcal{V}}|\psi|^2\,rdr=1,
\end{equation}
where $\mathcal{V}$ is the entire spatial domain. The Thomas-Fermi approximation is now updated to be given by 
\begin{equation}
	\label{tf_tor_sc}
	\psi_{\textsc{tf}}(r)=\left(\frac{\mu-A\exp(-l^2r^2)-\frac{1}{2}r^2-\frac{1}{r^2}}{g}\right)^{1/2},
\end{equation}
while the modified energy functional is
\begin{eqnarray}
	\label{e_tor_sc}
	E_f&=&\frac{1}{2}\int_\mathcal{V}|\nabla\psi|^2+\left(V(r)-\mu+\frac{1}{r^2}\right)|\psi|^2+\nonumber\\
	&\phantom{=}&\quad\frac{{\rm{i}}}{r^2}\left(\psi^*\frac{\partial\psi}{\partial\theta}-
	\psi\frac{\partial\psi^*}{\partial\theta}\right)+\frac{g}{2}|\psi|^4\, rdrd\theta,
\end{eqnarray}
which can be obtained from the transformation $\psi\rightarrow\psi\exp({\rm{i}}\theta)$ in (\ref{e_tor}). The energy is then obtained from $E=E_f-E_g$, where $E_g$ is the ground state energy and is evaluated as $E_g=E_f|_{\psi_0}$. Note that as expected
\begin{equation}
2{\rm i}\frac{\partial\psi}{\partial t}=\frac{\delta E_f}{\delta\psi^*}.
\end{equation}

The form of the angular momentum, however, is not so clear. Applying the transformation $\psi\rightarrow\psi\exp({\rm{i}}\theta)$ to (\ref{p_tor}) will give
\begin{equation}
	\label{p1_tor_sc}
	p_1=\frac{{\rm i}}{2}\int_\mathcal{V}(\psi-\psi_0)\frac{\partial\psi^*}{\partial \theta}-(\psi^*-\psi_0^*)\frac{\partial\psi}{\partial\theta}-|\psi-\psi_0|^2\, drd\theta,
\end{equation}
for ground state $\psi_0$. However this particular form of the angular momentum will be rejected and instead the following form (note its similarity to (\ref{p_tor})) of angular momentum will be used
\begin{equation}
	\label{p_tor_sc}
	p=\frac{{\rm i}}{2}\int_\mathcal{V}(\psi-\psi_0)\frac{\partial\psi^*}{\partial \theta}-(\psi^*-\psi_0^*)\frac{\partial\psi}{\partial\theta}\, drd\theta.
\end{equation}
The reason for the rejection of (\ref{p1_tor_sc}) is to ensure that the angular momentum of the elementary excitations are found and not the angular momentum of the elementary excitations plus the angular momentum caused by the persistent flow. Additionally it allows the group relationship $\Omega=\partial E/\partial p$ to be satisfied for energy given by (\ref{e_tor_sc}) and angular momentum given by (\ref{p_tor_sc}). Finally, an alternative form for the energy can be obtained by taking $\theta\rightarrow b\theta$ for constant $b$ in the expressions for energy and angular momentum (\ref{e_tor_sc}) and (\ref{p_tor_sc}) and considering the variational relationship (\ref{grp})
to give
\begin{equation}
	E=\int_{\mathcal{V}}\frac{1}{r}\left[\left|\frac{\partial\psi}{\partial\theta}\right|^2+
	\frac{{\rm i}}{2}\left(\psi^*\frac{\partial\psi}{\partial\theta}-\psi\frac{\partial\psi^*}{\partial\theta}\right)\right]\, drd\theta.
\end{equation}

The dynamics of solitary wave solutions in a toroidal trap with a persistent flow have been simulated for a range of values of interaction strength $30<g<750$ and parameters $A$ and $l$. The results for two typical examples are detailed below for $g=30$ and $g=150$. As before, in each case the energy-angular momentum dispersion curve is symmetric about $p=0$. The same colour scheme used in Sect.\ III will again be employed here. 

\subsection*{A. Interaction strength $g=30$ }
\label{toroidal_sc_30}

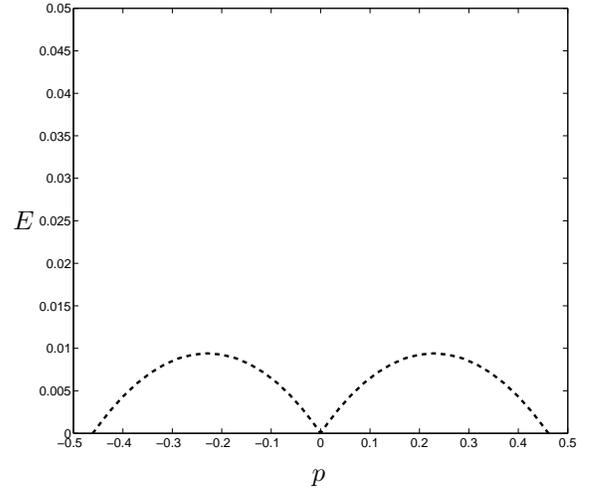
\begin{figure}[ht]
\centering
\includegraphics[scale=0.3]{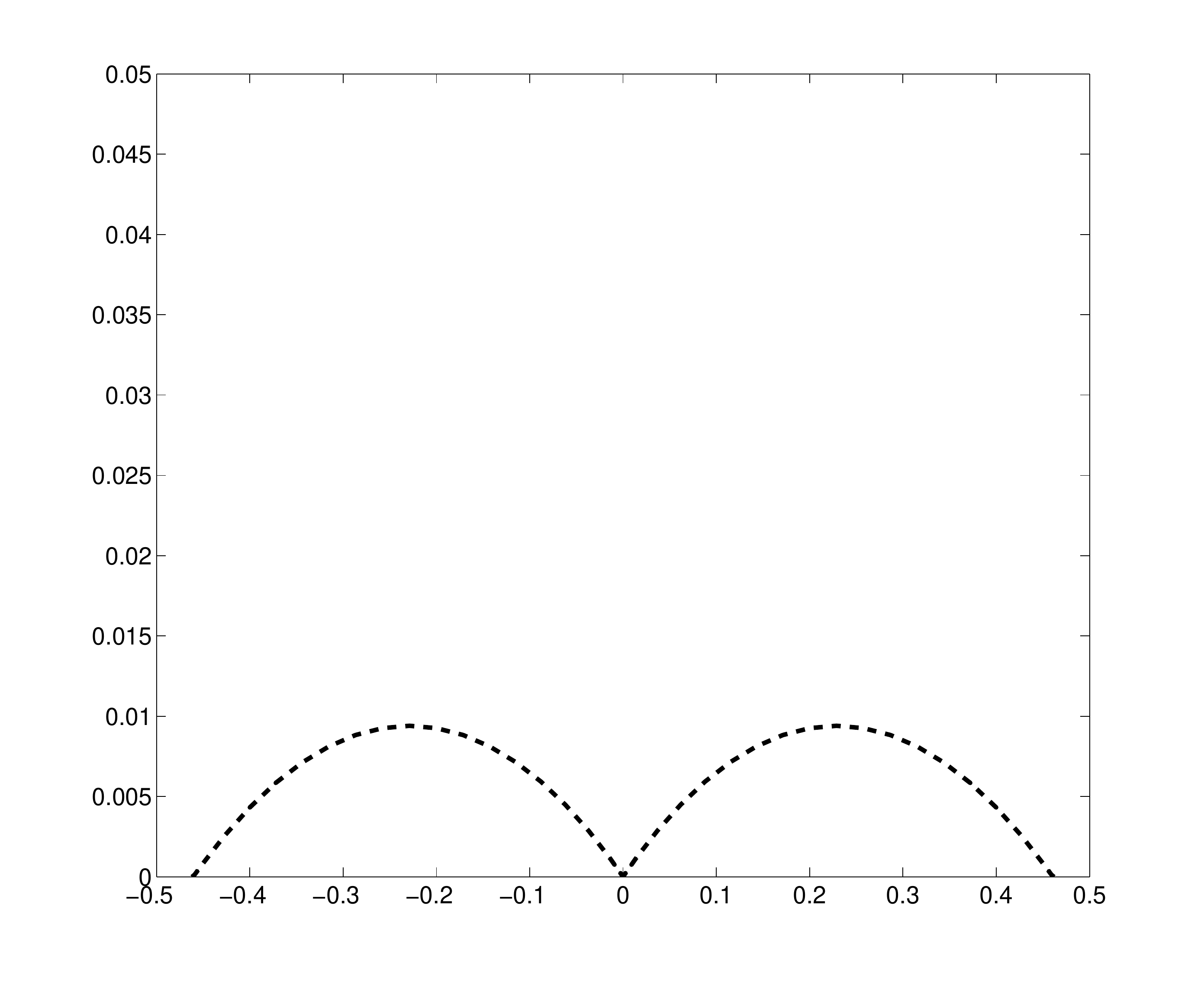}\\
\begin{picture}(0,0)(10,10)
\put(11,26) {$p$}
\put(-102,121) {$E$}
\end{picture}
\caption{\baselineskip=10pt \footnotesize The energy-angular momentum dispersion curve for the annular condensate with persistent flow governed by the GP equation (\ref{gp_tor_sc}) for parameter set $\{g, A, l\}=\{30, 15, 0.5\}$. Rarefaction waves are represented by the thick black dashed line. The dispersion curve is symmetric about $p=0$.}
\label{tor_sc_30_ep}
\end{figure}

The first case study will focus on a parameter range with a low interaction strength: $\{g,A,l\}=\{30,15,0.5\}$ with the corresponding value of the chemical potential $\mu=8.24$. The complete family of solitary wave solutions in the annulus are then found and the energy-angular momentum dispersion curve is shown in Fig.\ \ref{tor_sc_30_ep}. For the particular choice of parameters there is only one distinct solitary wave solution which exists in the range $-0.08\le\Omega\le0.09$ and is comprised solely of a single rarefaction wave along $\theta=\pi/2$ (see the thick black dashed line in Fig. \ref{tor_sc_30_ep}). At no energy or angular momentum does the rarefaction wave develop vorticity.

\subsection*{B. Interaction strength $g=150$}
\label{toroidal_sc_150}

\begin{figure}[ht]
\centering
\includegraphics[scale=0.3]{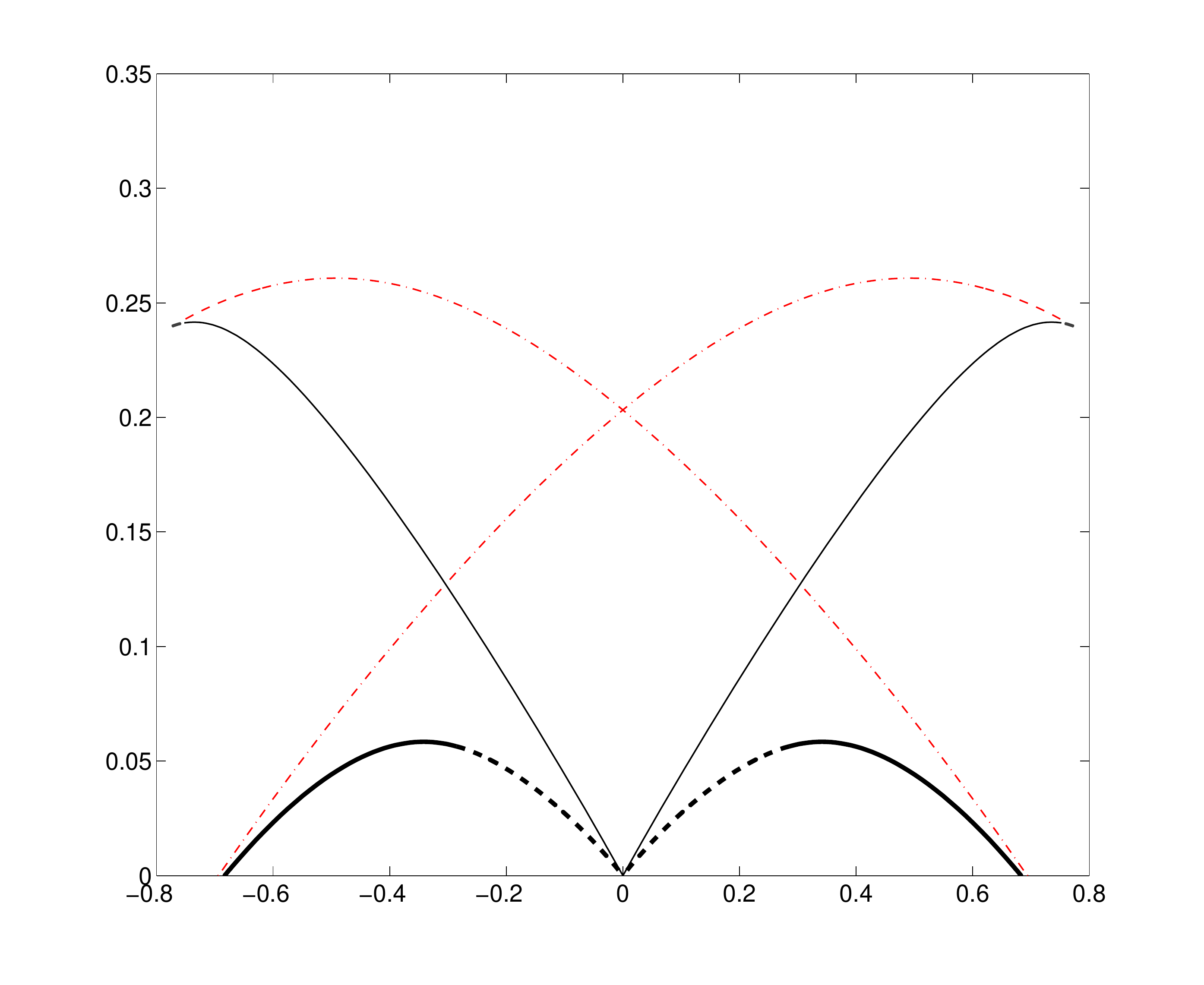}\\
\begin{picture}(0,0)(10,10)
\put(11,26) {$p$}
\put(-102,121) {$E$}
\end{picture}
\caption{\baselineskip=10pt \footnotesize (Color online) The energy-angular momentum dispersion curve for the annular condensate with persistent flow governed by the GP equation (\ref{gp_tor_sc}) for parameter set $\{g, A, l\}=\{150, 40, 1\}$. There exist two distinct single vortex solutions: thick solid black and thin solid black lines with corresponding rarefaction waves (dashed thick black and thin black lines respectively). In addition there is a two vortex solution (thin dashed-dot dark gray (red) line) with corresponding rarefaction wave (thin dashed dark gray (red) line). The dispersion curve is symmetric about $p=0$.}
\label{tor_sc_150_ep}
\end{figure}
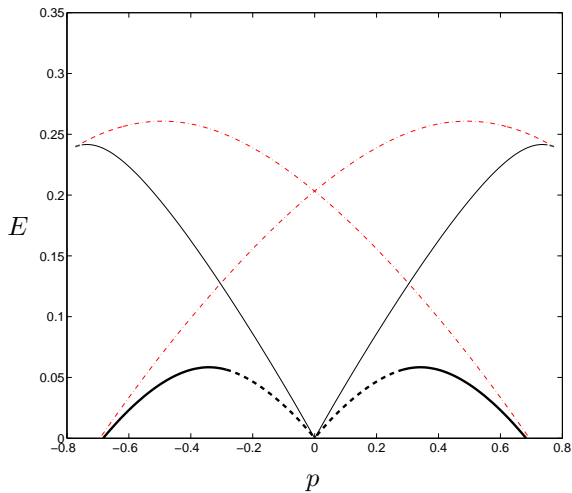

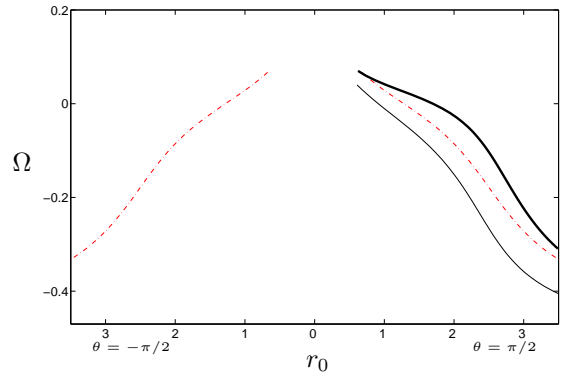
\begin{figure}[ht]
\centering
\includegraphics[scale=0.3]{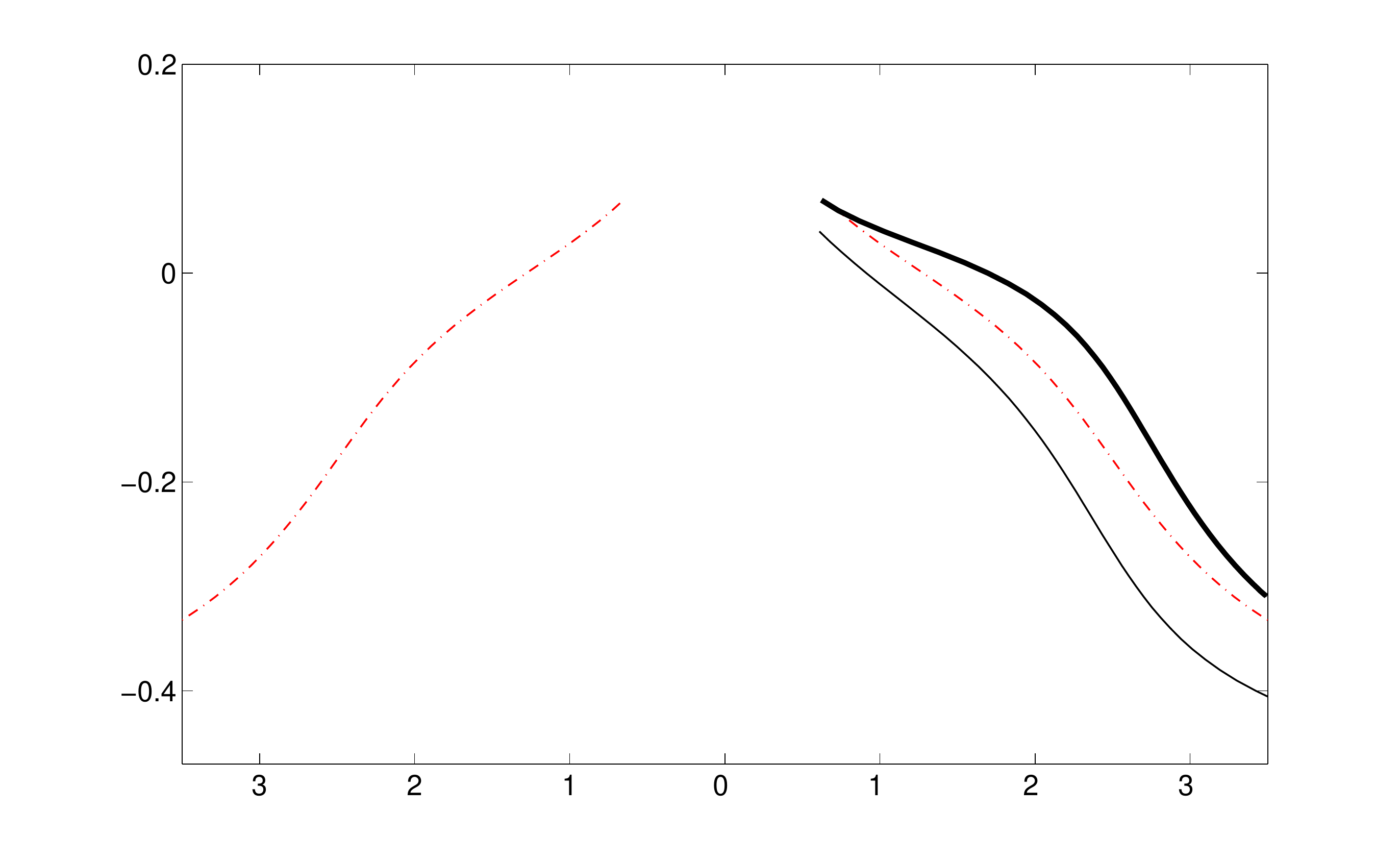}\\
\begin{picture}(0,0)(10,10)
\put(11,22) {$r_0$}
\put(-100,96) {$\Omega$}
\put(-70,28) {{\tiny{$\theta=-\pi/2$}}}
\put(74,28) {{\tiny{$\theta=\pi/2$}}}
\end{picture}
\caption{\baselineskip=10pt \footnotesize (Color online) The angular velocity against distance of the three different vortex solutions of $\{g, A, l\}=\{150, 40, 1\}$. A slice along $\theta=\pm\pi/2$ is taken. Colors refer to the different branches on the dispersion curve (Fig. \ref{tor_sc_150_ep}).}
\label{tor_sc_150_ud}
\end{figure}

The dispersion curve for $\{g,A,l\}=\{30,15,0.5\}$ was a typical example of the dynamics of the condensate in a persistent flow for low interaction strength. By increasing the interaction strength, a wider range of dynamical solutions will be exhibited. The second choice of parameter set is chosen to be $\{g, A, l\}=\{150, 40, 1\}$ with chemical potential $\mu=8.75$. Note that this is the same parameter set as chosen in Sect.\ IIIA, however the chemical potentials are different. For the parameter set there exist three distinct solitary wave solutions which can be seen in the energy-angular momentum dispersion curve plotted in Fig.\ \ref{tor_sc_150_ep}. A plot of the angular velocity of the vortex solutions against distance from the origin is given in Fig.\ \ref{tor_sc_150_ud}.

To describe the dynamics exhibited consider first the point $(E,p)=(0,0)$ where no solutions exist. If the energy and angular momentum are increased (to positive values) then a rarefaction wave develops along $\theta=+\pi/2$ and exists in the range of angular velocity $0.08\le\Omega\le0.33$. At $\Omega=0.08$ the rarefaction wave develops vorticity such that the solution is now a single vortex of positive circulation. The single vortex solution exists until $\Omega=-0.31$ at which point $(E,p)=(0,0.68)$ and the solution terminates. The rarefaction wave and the single vortex solution just described form the first branch of the dispersion curve. 

A cusp develops at $(E,p)=(0,0.68)$ and a new, second, branch is formed epitomised by the onset of two vortices of opposite circulation at the edge of the condensate, one along $\theta=+\pi/2$ and the other along $\theta=-\pi/2$ creating an anti-symmetric two vortex solution. The dynamics over the cusp are continuous (see Fig. \ref{tor_sc_150_ud}). The new branch that is formed exists for $-0.31\le\Omega\le0.15$ and is subdivided into two regions; the first region where the two vortex solution exists ($-0.31\le\Omega\le0.06$) and the second region where the two vortices lose circulation and become a pair of rarefaction waves ($0.07\le\Omega\le0.15$). 

The second branch terminates at $\Omega=0.15$ at which point a cusp develops over which the dynamics are again continuous. Note that the continuity of the dynamics cannot be seen in Fig.\ \ref{tor_sc_150_ud} since at the cusp the solutions are rarefaction waves. Tracing out the new third branch in the dispersion curve gives a system that contains a single rarefaction wave that exists in the range $0.04\le\Omega\le0.15$. At $\Omega=0.04$ the rarefaction wave develops vorticity and the solution is thus a single vortex which is present all the way until the branch terminates at $(E,p)=(0,0)$.

The dispersion curve for the parameter range chosen here therefore contains two distinct single solitary wave solutions (thick black and thin black curves in Fig. \ref{tor_sc_150_ep}) as was previously seen in annular condensates with no persistent flow present (Sect.\ III). Note however that at $(E,p)=(0,0)$ where the branches of the two single solitary wave solutions meet the dynamics are not continuous, as can be seen in Fig.\ \ref{tor_sc_150_ep} where the first branch contains a rarefaction wave (thick black dashed line) and the third branch which contains a single vortex solution (thin black solid line). 

The two case studies ($\{g,A,l\}=\{30,15,0.5\}$ and $\{150,40,1\}$) have concentrated on low-to-mid range interaction strengths. For higher $g$ the energy-angular momentum dispersion curves become too complicated to be described in a detailed yet concise fashion. However it can be appreciated that the net result of increasing $g$ will be to produce new distinct solitary wave solutions and thus many branches in the dispersion curve. A further result of increasing $g$ will be in the decrease in the range of angular velocities for which the rarefaction waves exist. 

\section*{VII. CONCLUSION}

This paper has investigated the dynamics of a two-dimensional
condensate held under a toroidal trapping potential. A toroidal
trapping potential creates an annular condensate  
  with richer and broader range of the solitary waves in comparison
  with the semi-infinite channel geometries \cite{mb,kp2}.

The toroidal condensate is dependent on three independent parameters
(that can be set experimentally) $\{g,A,l\}$, of which the interaction
strength $g$ is the dominant parameter. Changing the value of $g$
determines the number of different types of solutions on the
dispersion curve. For low values of $g$ there are two fundamental
branches of solutions (see Fig.\ \ref{tor_150_ep} for
$\{g,A,l\}=\{150,40,1\}$). For a larger value of $g$, the number of
fundamental branches of solutions increases. The parameter set
$\{g,A,l\}=\{500,100,0.9\}$ has been detailed extensively in this
paper. There are seven fundamental branches of solutions (see Fig.\
\ref{tor_500_ep}) which contain up to four vortices in the
condensate. An interesting scenario arises in which there exists two
different branches each of which contains a single vortex with the
vortices of opposite circulation. There is a solution where the two
branches possess a vortex with the same angular velocity and are at
the same distance from the origin. Since the vortices have opposite
circulation, such a result seems rather surprising.
One of the solutions was determined to contain not just the vortex,
but in addition rarefaction waves embedded in the background of the condensate which resulted from the decay of a vortex from a different branch on the dispersion curve.

The stability of the seven fundamental branches of the parameter set $\{500,100,0.9\}$ was also investigated by evolving the solitary wave solutions forward in time. The solutions were evolved until three of four complete revolutions were completed, or the solutions decayed into sound waves. It was determined numerically that, if the solitary wave sequence contains more than one vortex for any given angle, then the branch is unstable. All the other branches were determined to be stable. 

By isolating the solitary waves from type I and type II solutions it is possible to investigate the possible range of collisional effects and compare them with a point vortex model. The solutions in this paper are all obtained by numerically simulating the Gross-Pitaevskii equation for the inhomogeneous condensate, whereas point vortex models consider a `box-like' condensate where the density is uniform in the annular region and is zero outside this region. It is seen in this paper that collision of the vortices result in two distinct behaviours: either the vortices collide elastically or collide inelastically. Elastic collisions occur for low angular velocities where the vortices repel each other or for high angular velocities where the vortices pass through one another. For intermediate angular velocities, the vortices collide inelastically and decay into sound waves. Point vortex models are only able to capture the collisions for low angular velocities.

Finally, a separate condensate is considered where now a persistent flow is included into the condensate. It is interesting to note the similarities and differences between the annular condensate without a persistent flow (Sect.\ III) and one with a persistent flow (Sect.\ VI). 

\section*{ACKNOWLEDGMENTS}

NGB acknowledges grant support from EPSRC.


\begin{thebibliography}{1}
\bibitem{num}{In fact whenever there is only a single solitary wave solution present in the condensate, the solitary wave is numerically computed at $\theta=\pi$. When more than one solitary wave is present in the condensate the solitary waves are taken numerically to be along $\theta=\pm\pi/2$. Therefore to agree with the future notation the single solitary wave is positioned at $\theta=\pi/2$.}
\end{thebibliography}
\end{document}